\documentclass[pra,twocolumn,amsmath,amssymb]{revtex4-1}            

\usepackage{xifthen}

\newcommand{\refAppendix}[6]{#1
  \ifthenelse{\isempty{#2}}%
    {}
    {\protect\cite{#2}}
    #3\protect\ref{#4}#5#6\xspace
}

\usepackage{forloop,ifthen} 
\usepackage{multirow} 
\usepackage{ esint }

\usepackage{amsmath} 
\usepackage{graphicx}
\usepackage{float}
\usepackage{transparent}
\usepackage{fancybox}
\usepackage{color}
\usepackage{dcolumn}
\usepackage{bm}
\newcommand{\VEC}[1]{{\mathbf #1}}
\usepackage{xspace}

\usepackage{hyperref}
\usepackage{calc}

\newcommand{\CO}[1]{}

\newcommand{\suppress}[1]{}

\newcommand{\squeezedHeight}{36.75mm}
\newcommand{\squeezedWidth}{108mm}
\renewcommand{\squeezedHeight}{33mm}
\renewcommand{\squeezedWidth}{87mm}

\usepackage[usenames,dvipsnames]{xcolor}
\usepackage{ amssymb }

\newcommand{\ps}{phase space\xspace}

\newcommand{\emphCaption}[1]{{\it{#1}}}
\newcommand{\emphLabelbracketed}[1]{\textcolor{black}{\textsf{\bf{(#1)}}}}
\newcommand{\emphLabel}[1]{\textcolor{black}{\textsf{{(#1)}}}}
\newcommand{\bea}[1]{\begin{eqnarray}\label{#1}}
\newcommand{\eea}{\end{eqnarray}}
\newcommand{\corrected}[2]{#2}

\begin{document}

\title{{Dynamic Shear Suppression in Quantum Phase Space}}

\author{Maxime Oliva and Ole Steuernagel}

\affiliation{School of Physics, Astronomy and Mathematics, University of
Hertfordshire, Hatfield, AL10 9AB, UK}

\date{\today}

\begin{abstract}
  Classical \ps flow is inviscid. Here we show that in quantum \ps Wigner's probability
  current~$\VEC{J}$ can be effectively viscous. This results in shear suppression in
  quantum \ps dynamics which enforces Zurek's limit for the minimum size scale of spotty
  structures that develop dynamically. Quantum shear suppression is given by gradients of
  the quantum terms of~$\VEC{J}$'s vorticity. Used as a new measure of quantum dynamics
  applied to several evolving closed conservative 1D bound state systems, we find that
  shear suppression explains the saturation at Zurek's scale limit and additionally
  singles out special quantum states.
  \\\\
  DOI: \href{https://doi.org/10.1103/PhysRevLett.122.020401}{10.1103/PhysRevLett.122.020401} 
\end{abstract}

\maketitle

The differences between quantum and classical evolution are best investigated in
\ps~\cite{Gong_Brumer_PRA03}.  It is known that quantum evolution in \ps does not obey
Liou\-ville's theorem of volume conservation~\cite{Moyal_MPCPS49,Oliva_PhysA17} and that
there is no velocity field in quantum \ps~\cite{Oliva_PhysA17} (and therefore no flow).
It is less clear why there is no quantum
chaos~\cite{Zurek_Paz_PRL94,Casati_PRL94_comment,Zurek_Paz_PRL94_reply,Gong_Brumer_PRA03,Gomez_CSF14}.

Here we show that the effective viscosity of quantum \ps dynamics entails a shear
suppression mecha\-nism that fundamentally differentiates quantum from classical
dynamics. This observation explains, amongst other things, the absence of quantum chaos.

\begin{figure}[b]
  \includegraphics[width=87mm]{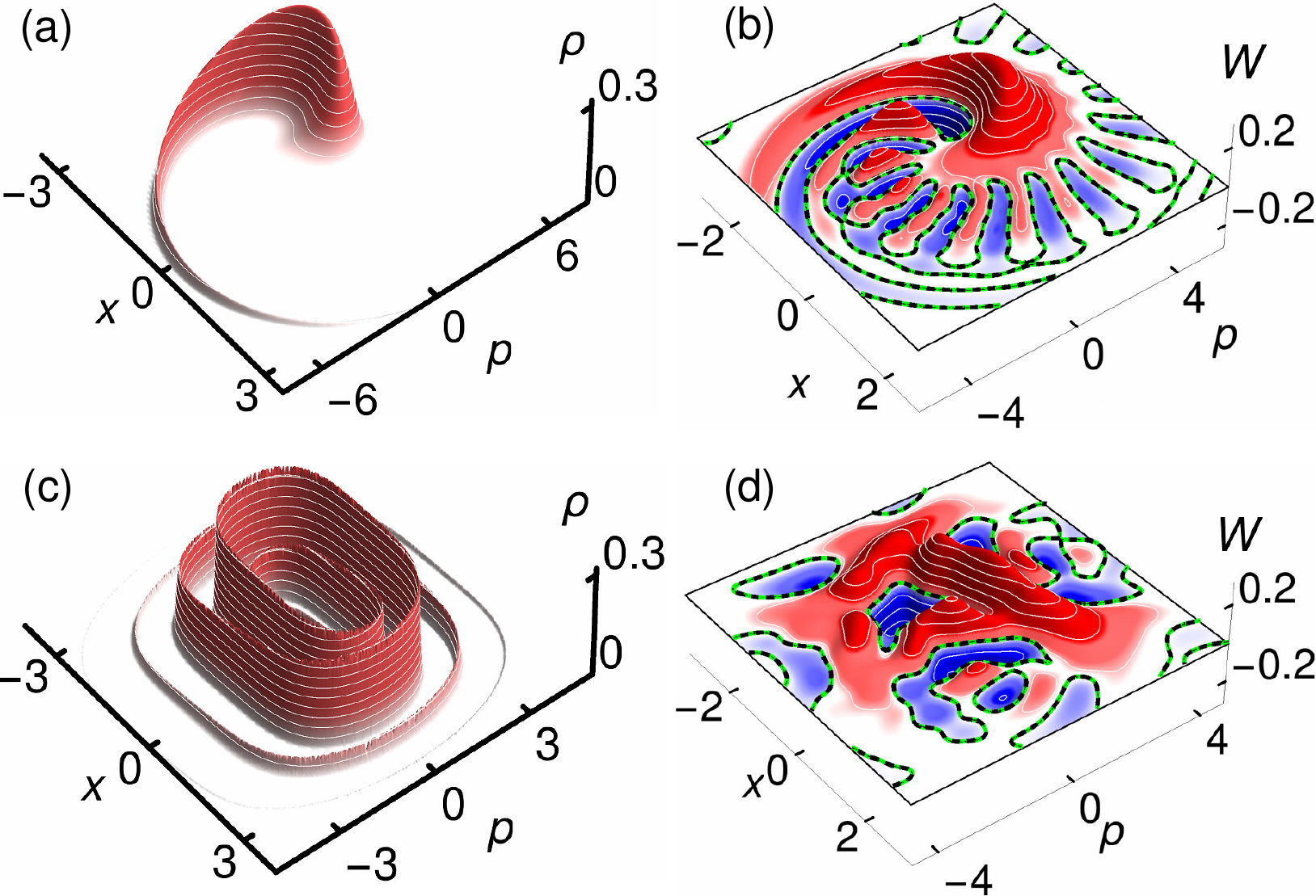}
  \caption{\CO{(Color online) }\emphCaption{Comparison between classical and quantum
      distributions in \ps.}  For short times quantum evolution resembles classical
    evolution, compare~\emphLabelbracketed{b} to~\emphLabelbracketed{a}. But for long times, since quantum
    evolution creates less fine structures than classical
    evolution~\protect{\cite{Berry_JPA79,Zurek_NAT01}}, their outcomes differ very substantially,
    contrast~\emphLabelbracketed{d} with~\emphLabelbracketed{c}. A weakly excited initial
    state~$W_0(x,p,1.5,0)$ is propagated in the soft
    potential~$V_{\sf V} = {31 x^2}/{10} - { x^4}/{81}$ for time $t=50$,
    under,~\emphLabelbracketed{a}, classical evolution and,~\emphLabelbracketed{b}, quantum evolution.
    Similarly, the state~$W_0(x,p,2,0)$ is propagated in the hard
    potential~$V_{\sf U}=(x/2)^4$ for time~$t=25$, under,~\emphLabelbracketed{c}, classical
    and,~\emphLabelbracketed{d}, quantum evolution.\label{fig:Distributions}}
\end{figure}

We consider dynamics in 1D closed conservative systems with spatial coordinate~$x$ and
momentum~$p$. As initial states we use displaced
Gaussians~$W_0(x,p,x_0,p_0)=(\pi \hbar)^{-1} \exp[{-(x-x_0)^{2} - {(p-p_0)^{2}}/{\hbar^{2}} }]$
which are positive~\cite{Hudson_RMP74} and therefore ``classical''; $\hbar = h / 2\pi$ is Planck's
constant.  After sufficient time~$t$ under classical evolution, such a nonsingular initial
distribution typically forms thinly stretched out threads, see
Fig.~\ref{fig:Distributions}~\emphLabelbracketed{c}. Generally, structures of a classical
probability distribution~$\rho(x,p,t)$ become progressively finer as time
progresses~\cite{Zurek_NAT01}, particularly chaotic systems develop very fine structures
quickly~\cite{Berry_AIPC78,Zurek_NAT01,Nolte_PT10,Cvitanovic_Chaos_book_12}.

Wigner's quantum \ps distribution $W(x,p,t)$ \cite{Wigner_PR32,Hillery_PR84} is the
closest quantum
analog~\cite{Hillery_PR84,Leibfried_PT98,Zurek_NAT01,Tilma_PRL16,Oliva_PhysA17} of the
classical \ps distribution~$\rho$.  Quantum evolution creates negative regions (blue,
deline\-ated by dashed lines at~$W=0$, see Fig.~\ref{fig:Distributions}) [in all figures
atomic units $\hbar=1$ and $M=1$ are used~\protect\refAppendix{ (see
  Appendix}{}{}{sec_UnitsVorticity}{)}{].} These negative regions represent the existence
of quantum coherences, see Refs.~\cite{Feynman_NegEssay87,Leibfried_PT98,Zurek_NAT01,Oliva_PhysA17}
and\refAppendix{ Appendix}{}{}{sec_StructureFormation}{}{.}

Interference in \ps~\cite{Schleich_01} is a property built into quantum \ps functions,
such as $W$, through the Wigner-Moyal mappings~\cite{Wigner_PR32,Moyal_MPCPS49} between
Hilbert space operators and their quantum \ps
images~\cite{Rasinariu_FP12,Hancock_EJP04}. This interference limits the fineness of
spotty structures, that $W$ can have, to Zurek's \ps area scale~\cite{Zurek_NAT01}
\begin{eqnarray}
a_Z = \frac{h}{P} \frac{h}{L} = \frac{2 \pi}{K_x}  \frac{2 \pi}{K_p} 
\label{eq:aZurek}
\end{eqnarray}
(see Fig.~\ref{fig:Distributions}~\emphLabelbracketed{b} and~\emphLabelbracketed{d}). Here
\corrected{$h$ is Planck's constant, and }{}length~$L$ and momentum~$P$ are~$W$'s spread in \ps and thus the area~$L\, P$ (measured in
units of action) to which it is confined. The maximal wave numbers associated with $W$'s
structures in $x$ and $p$ are, respectively, $K_x = P/\hbar$ and
$K_p = L/\hbar$~\cite{Zurek_NAT01} (see Ref.~\cite{Oliva_superoscillations_PRA17} for
exceptions). Over time states develop spotty structures that saturate on the Zurek
scale~$a_Z$~\cite{Zurek_NAT01}.

Here we show that the adherence to Zurek's scale limit in the evolution is best understood
in terms of the visco\-sity\refAppendix{, see Appendix}{}{}{sec_Viscosity}{,}{ of} the
Wigner current $\VEC{J}$~\cite{Wigner_PR32,Ole_PRL13,Kakofengitis_PRA17,Oliva_PhysA17}.

$\VEC{J}$ is the quantum analog of the classical \ps current~$\VEC{j} = \rho \VEC{v}$
which transports the probability density~$\rho(x,p,t)$ according to Liou\-ville's continuity
equation~$\partial_t \rho = - {\bm \nabla} \cdot \VEC{j}$. Here~$\VEC{v}$ is the classical
\ps velocity $ \VEC{v}=[p/M,-\partial_x V(x)]$,~$M$ the mass of the particle, $V(x)$
the potential\corrected{}{, and ${\bm \nabla} = (\partial_x,\partial_p) $ the gradient operator}.

\begin{figure}[t]
\includegraphics[width=0.995\columnwidth]{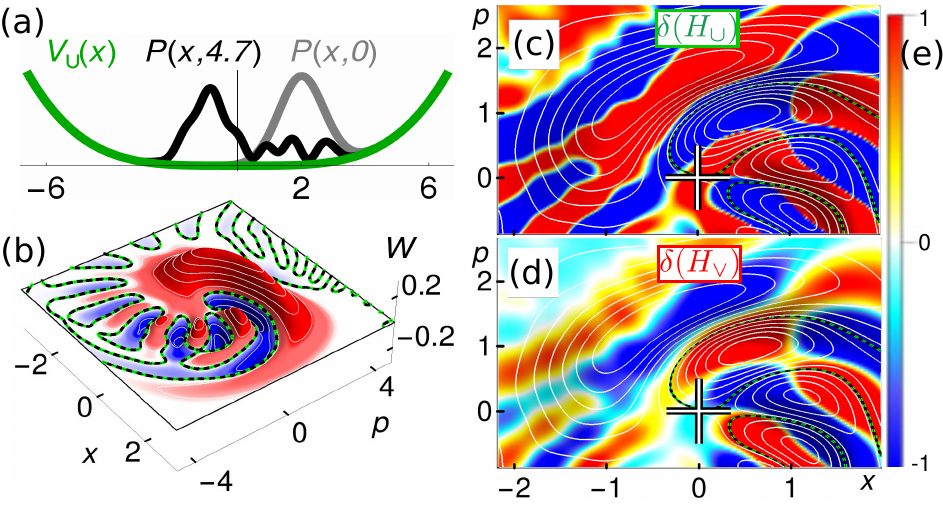}
\caption{\CO{(Color online)} \emphCaption{Polarization of the vorticity $\delta$ and
    inversion of this polarization.}  The comparison between~\emphLabelbracketed{c}
  and~\emphLabelbracketed{d} shows polarization inversion.  \emphLabelbracketed{a}, sketch of hard potential
  $V_{\sf U}=(x/2)^4$ together with probability distributions~$P(x,t)=|\Psi(x,t)|^2$
  (black curve) of state evolved in~$V_{\sf U}$ from initial state~$W_0(x,p,2,0)$ (grey
  curve).~$W$ for~$t =4.7$ in \emphLabelbracketed{a} is shown in \emphLabelbracketed{b}. White contours of~$W$ [the
  origin~$(x,p)=(0,0)$ is labeled by a white cross] are overlaid with colors [legend
  given in sidebar \emphLabelbracketed{e}] representing values of
  tanh[$50 \; \delta(H_{\sf U})$],~\emphLabelbracketed{c}, and
  tanh[$50 \; \delta(H_{\sf V})$],~\emphLabelbracketed{d}. For the Hamiltonians $H_{\sf V}$ and
  $H_{\sf U}$ the same potentials as in Fig.~\ref{fig:Distributions} are used.
  \label{fig:LocalShear}}
\end{figure}

Over time,~$\rho$ gets sheared since~$\VEC{v}$ creates nonzero gradients of its angular velocity
across energy shells. The classical Hamiltonian \ps flow is inviscid as $\VEC{v}$ is independent
of~$\rho$. Thus no terms suppress the effects of the gradients of the angular velocity, and so, as
time progresses, nonsingular probability distributions in \ps get sheared into ever finer filaments
[see~Fig~\ref{fig:Distributions}~\emphLabelbracketed{c}].

We define classical \ps shear as\refAppendix{ (see Appendix~}{}{}{sec_UnitsVorticity}{)}{} 
\begin{flalign}
  s (x,p;H) = \partial_{\widehat {{\bm \nabla}}_{\!\!H}} ( - {\bm \nabla} \times  \VEC{v} ) =
\partial_{\widehat {{\bm \nabla}}_{\!\!H}} (  \partial_p v_x - \partial_x v_p) \; , 
\label{eq:_ClassicalShear}
\end{flalign}
using the directional derivative across energy
shells~$\partial_{\widehat {{\bm \nabla}}_{\!\!H}}$, formed from the normalized gradient
$\widehat {\bm \nabla}_{\!\!H} = {\bm \nabla} H/|{\bm \nabla} H|$ of the
Hamiltonian~$H = \frac{p^2}{2M} +V(x)$.

The sign convention with the negative curl
in~$s$ in Eq.~(\ref{eq:_ClassicalShear}) was chosen to yield a positive sign for
\emph{clockwise} orientated fields since this is the prevailing direction of the classical
velocity field~$\VEC{v}$.  This choice yields
$s>0$ for hard potentials (potentials for which the magnitude of the force increases with
increasing amplitude), since they induce \emph{clockwise} shear,
see~Fig.~\ref{fig:Distributions}~\emphLabelbracketed{c}.
$s=0$ for harmonic oscillators and free particles, and
$s<0$ for soft potentials (potentials for which the magnitude of the force decreases with
increasing amplitude), since they induce \emph{anticlockwise} shear,
see~Fig.~\ref{fig:Distributions}~\emphLabelbracketed{a}.

\corrected{The quantum continuity 
equation is}{ $W$'s evolution is governed by the quantum continuity 
equation}~\cite{Wigner_PR32,Oliva_PhysA17}
\begin{equation} 
\partial_t W = - {\bm \nabla} \cdot \VEC{J} = - \partial_x J_x - \partial_p J_p .
\label{eq:W_Continuity}
\end{equation} 
\corrected{where ${\bm \nabla} = (\partial_x,\partial_p) $ is the gradient operator.}
Wigner's current~$\VEC{J}$ does not factorise
like~$\VEC{j} = \rho \VEC{v}$~\cite{Oliva_PhysA17}. It has an integral
representation\refAppendix{, see~}{Wigner_PR32,Kakofengitis_PRA17}{ and
  Appendix~}{sec_W_Special}{}{.} If the potential~$V(x)$ is smooth such that it can be
expanded into a Taylor series, the integral for $\VEC{J}$ can be determined explicitly
as~\cite{Wigner_PR32,Groenewold_Phys46,Moyal_MPCPS49}
\begin{flalign} {\VEC{J}(x,p,t)} & = \VEC{j} + \VEC{J}^Q
  = W  \VEC{v} 
  + \genfrac{(}{)}{0pt}{}{0}{J_p - j_p}
  \\
 = \! W & \! \genfrac{(}{)}{0pt}{}{ \frac{p}{M}}{- \partial_x V}
  + \left(
    \genfrac{}{}{0pt}{}{ 0
    }{-\sum\limits_{l=1}^{\infty}{\frac{(i\hbar/2)^{2l}}{(2l+1)!}  \partial_p^{2l} W
        \partial_x^{2l+1} V }} \right).
\label{eq:CurrentComponents}
\end{flalign}

$\VEC{J}$'s zeroth-order term in~$l$ is the classical term~$\VEC{j} = W \VEC{v}$
(e.g. $J_{p}|_{l=0} = - W \partial_x V$).  Terms of order $l\geq 1$ are the quantum
correction terms~$\VEC{J}^Q=\VEC{J}-\VEC{j}$. They are only present for anharmonic
potentials~\cite{Kakofengitis_PRA17}, which is why only anharmonic potentials create
coherences. Harmonic systems' \ps dynamics is
classical\refAppendix{, see Refs.}{Oliva_PhysA17,Kakofengitis_PRA17}{ and
  Appendix~}{sec_StructureFormation}{}{.}

The reaction of quantum dynamics to classical shear~$s$ has to reside in $\VEC{J}^Q$. To
extract it we form the vorticity $\delta$ of~$\VEC{J}^Q$
\begin{flalign}
  \delta (x,p,t;H) = - {\bm \nabla} \times  \VEC{J}^Q = \partial_p J_x^Q - \partial_x J_p^Q \; . 
\label{eq:shear_deviation}
\end{flalign}
$\delta$'s sign distribution shows a pronounced polarization pattern, see for example
Fig.~\ref{fig:LocalShear}~\emphLabelbracketed{c}: specifically, on the positive main ridge of~$W$
(Fig.~\ref{fig:LocalShear}~\emphLabelbracketed{b}) $\delta$ tends to be posi\-tive on the inside
(towards the origin) and negative on the outside. Because of this, the outside is being
slowed down while the inside speeds up. This polarized distribution of~$\delta$ therefore
counteracts the classical shear~($s_{V_{\sf U}}>0$) and can suppress it altogether.

The same applies to other positive regions of~$W$, whereas for its negative regions the
current~$\VEC{J}$ tends to be inverted~\cite{Ole_PRL13,Kakofengitis_EPJP17} just
as~$\delta$'s polarization pattern.

Switching the governing potential from hard,~$V_{\sf U}$, to soft,~$V_{\sf V}$ (using the
same state but different dynamics), reverses the classical shear, see
Fig.~\ref{fig:Distributions}. Accordingly, a reversal of the polarization pattern of
Fig.~\ref{fig:LocalShear}~\emphLabelbracketed{c} occurs in Fig.~\ref{fig:LocalShear}~\emphLabelbracketed{d}.

The distribution of~$\delta$'s polarization can be picked up with the directional
derivative~$\partial_{\widehat {{\bm \nabla}}_{\!\!H}} \delta(t;H)$. This we multiply
with~$W$, because nega\-tive regions of~$W$ invert the current~$\VEC{J}$~\cite{Ole_PRL13},
and because we want to weight it with the local contribution of the state. The resulting
measure for weighted shear polarization is
${\pi}(x,p,t;H) = W(t) \; \partial_{\widehat {{\bm \nabla}}_{\!\!H}} \delta(t;H)$.  Its
average across \ps is $W$'s shear polarization
\begin{flalign}
  {\Pi}(t;H) = \langle \!\!\,\langle {\pi}(t;H) \rangle \!\!\, \rangle =
  \iint_{-\infty}^{\infty} dx dp \; \pi(x,p) \, .
\label{eq:effective_viscosity_quantification}
\end{flalign}

\begin{figure}[t]
\centering
  \includegraphics[width=0.49\columnwidth]{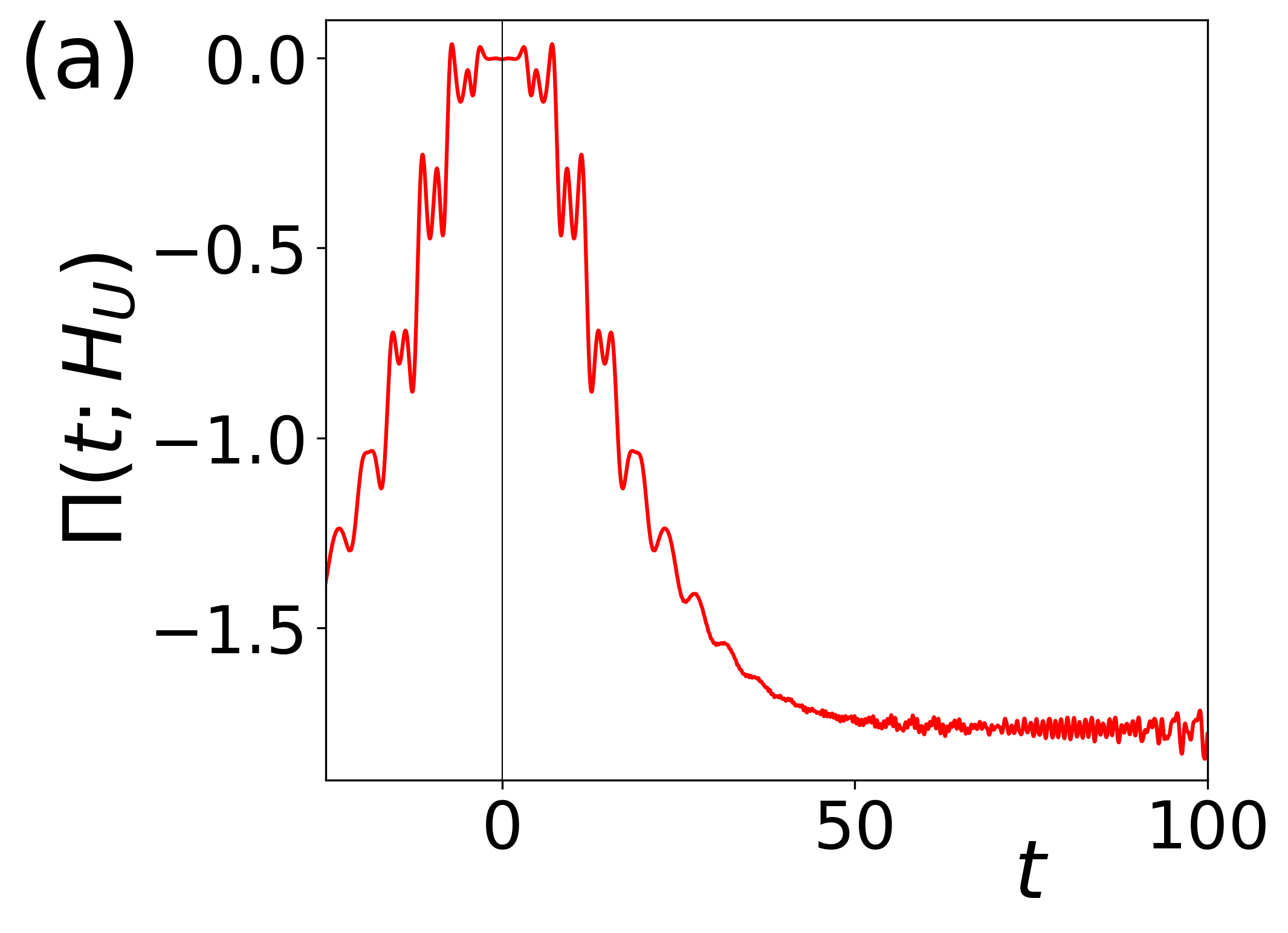}
  \includegraphics[width=0.49\columnwidth]{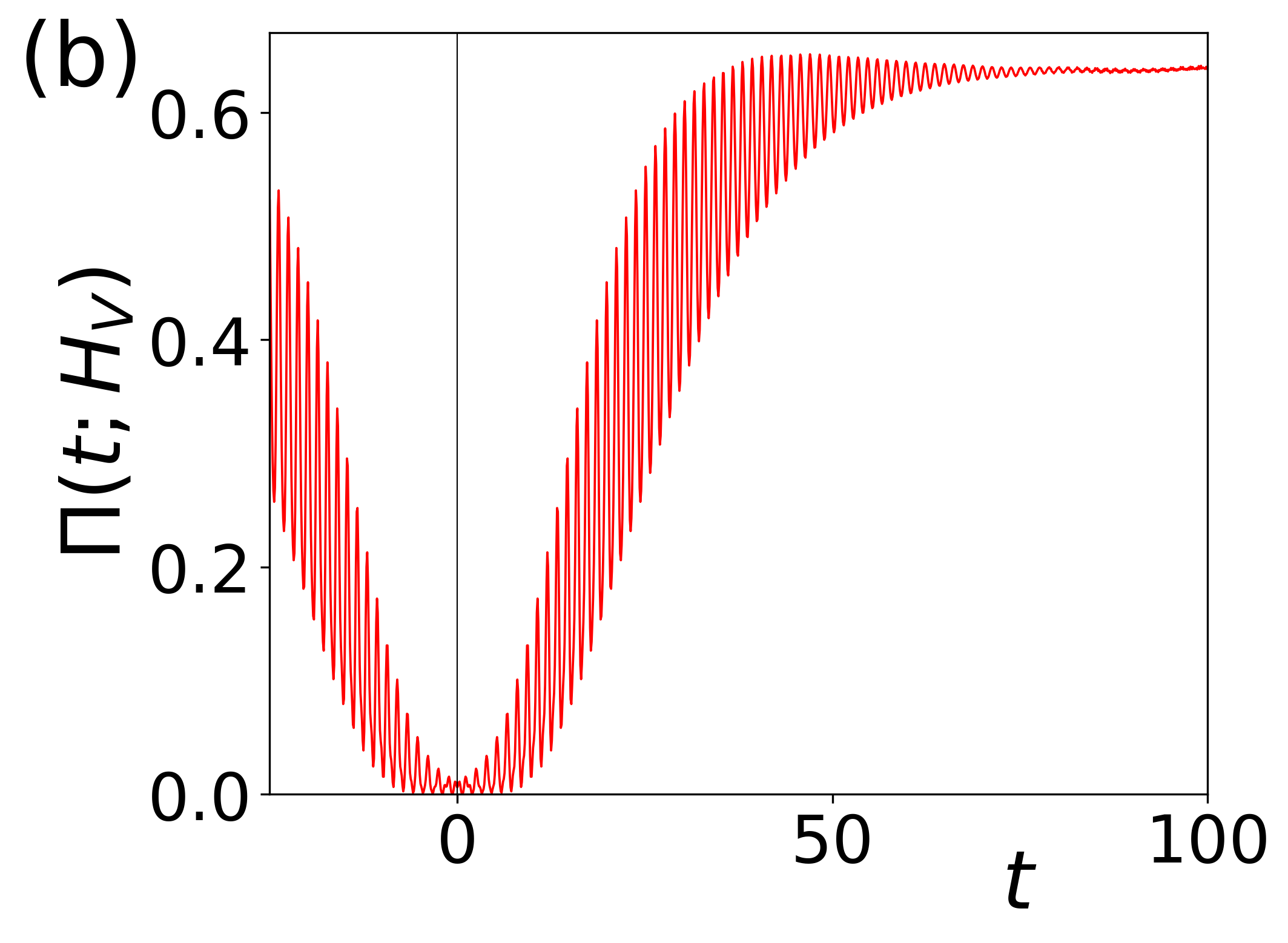}
  \caption{\emph{$\Pi(t;H)$ levels off over time as systems saturate.} $\Pi$'s time
    evolution for,\textcolor{black}{~\emphLabelbracketed{a}, initial state~$W_0(x,p,9,0)$, for the
      hard potential~$V_{\sf U}=x^4/500$, and,~\emphLabelbracketed{b}, $W_0(x,p,3,0)$, for the soft
      potential~$V_{\sf V}={31 x^2}/{10} - { x^4}/{81}$.  In accord with our
      sign-convention for Eq.~(\ref{eq:_ClassicalShear}) $\Pi(t;H_{\sf U})$ drops over
      time whereas $\Pi(t;H_{\sf V})$ rises, until the system saturates.}}
  \label{fig:GlobalShear}
\end{figure}

Initially $|\Pi(t)|$ rises on average and after a while levels off and settles, see
Fig.~\ref{fig:GlobalShear}.

We emphasize that the leveling-off behaviour of~$\Pi(t)$ is in marked contrast to the
classical case: in simple bound state systems the states never saturate, instead, for long
enough times
$ \langle \!\!\,\langle \partial_{\widehat {{\bm \nabla}}_{\!\!H}} (- \VEC{\nabla} \times
\VEC{j}) \rangle \!\!\, \rangle \propto t$ since $\rho(t)$ gets stretched out linearly
into ever finer threads, see Fig.~\ref{fig:Distributions}~\emphLabelbracketed{c}
and\refAppendix{ Appendix~}{}{}{subsubsec:Appendix_sCaseDistinctionBreakdown}{}{.} Also the
quantum evolution can shrink structures of~$W$ in size, but~$W$'s mini\-mal structures are
forced to saturate at the Zurek scale by shear suppression.

When a state saturates, the gradients in the quantum terms of~$\VEC{J}$ become so large
that they strongly quantum suppress the classical shear inherent in~$\VEC{J}$.  Where
mini\-mal structures of~$W$ have formed, this quantum shear suppression prevents still
finer structures from developing: $\VEC{J}$'s effective viscosity enforces the
saturation of states at the Zurek scale.

\begin{figure}[t!]
\centering
  \includegraphics[width=0.99\columnwidth]{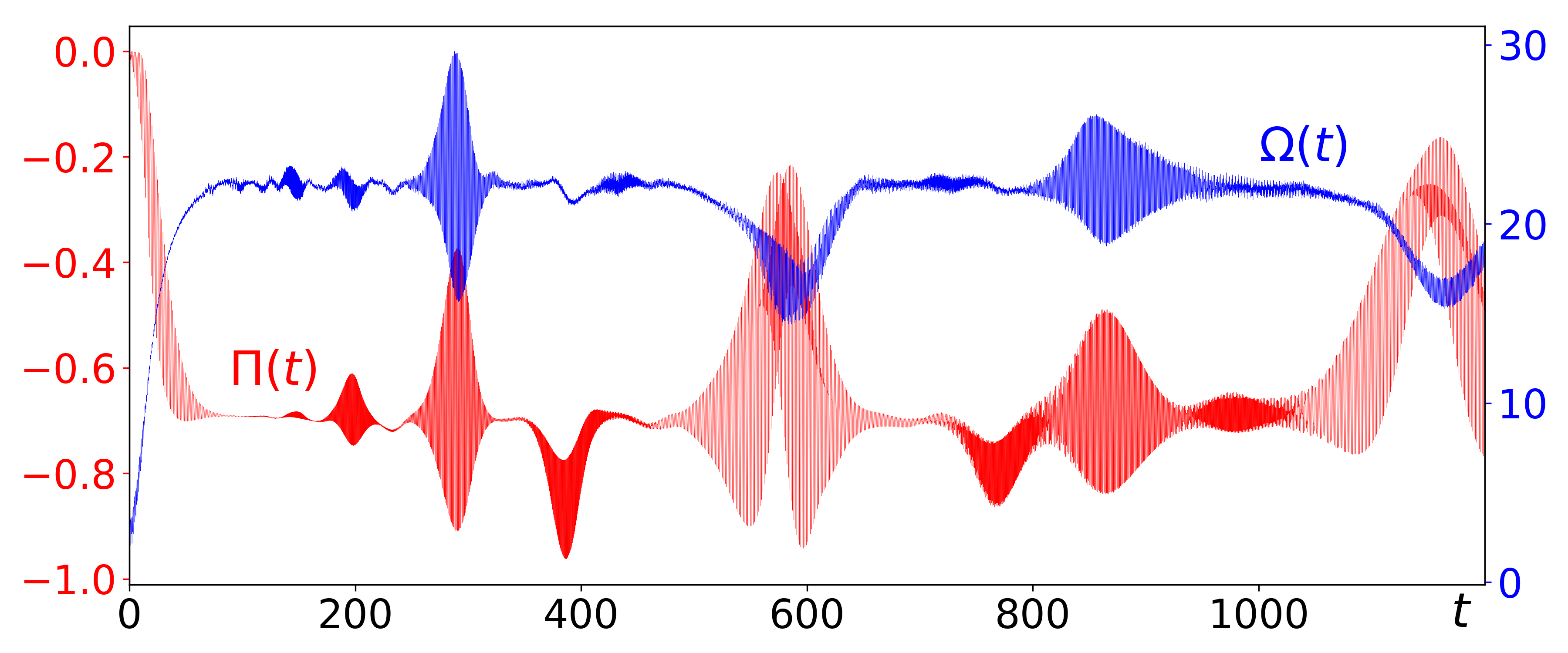}
  \caption{\emphCaption{Shear suppression $\Pi(t)$ and frequency contents $\Omega(t)$ can
      evolve similarly.} A Gaussian state~$W_0(x,p,3,0)$ is evolved in a hard
    potential~$V_{\sf U} ={31 x^2}/{10} + { x^4}/{81}$. While $\Omega(t)$ rises (here,
    $\Omega_\text{max} = 182.5$), as~$W$ develops fine structure, $\Pi(t)$ drops, since we
    consider a hard potential~$V_{\sf U}$. Both curves~$\Pi(t)$ and~$\Omega(t)$ level off
    at the time where $W$'s fine structures saturate at the Zurek scale. Here the
    oscillations around the respective mean values for saturated systems are due to the
    formation of special (partial revival)
    states~\protect{\cite{Averbukh_PLA89,Robinett_PR04}}, for details see
    Fig.~\ref{fig_Soft_Tableau} and\protect\refAppendix{
      Appendix~}{}{}{sec:Appendix_SpecialStates}{}{.}\label{fig_Pi_vs_Omega}}
\end{figure}

When this happens $\Pi(t)$ has settled, see
Figs.~\ref{fig:GlobalShear},~\ref{fig_Pi_vs_Omega} and~\ref{fig_Soft_Tableau}.

To make explicit the connection between shear suppression $\Pi(t)$ and the saturation of
systems at the Zurek scale we define~$W$'s spatial frequency contents~$\Omega$ as
\vspace{-0.1cm}
\begin{flalign} \label{eq:Omega_W_FrequencyContent} \Omega(t) = \frac{\iint dk_x dk_p \;
    |\overset{\approx}{W}(k_x,k_p,t) \; k_x k_p|}{\iint dk_x dk_p \;
    |\overset{\approx}{W}(k_x,k_p,t)|} < 2 K_X K_P ,
  \end{flalign}
  
\vspace{-0.2cm}
\noindent
where~$\overset{\approx}{W}(k_x,k_p)$ is the 2D Fourier transform of~$W(x,p)$.  Since a
state cannot only consist of structures at the Zurek scale
Eq.~(\ref{eq:Omega_W_FrequencyContent}) obeys the
inequality~$\Omega < \Omega_\text{max} = \frac{8 \pi^2}{a_z} = 2 K_X K_P$, compare
Eq.~(\ref{eq:aZurek}) and\refAppendix{ Appendix~}{}{}{sec:Appendix_OmegaLimit}{}{.}

Figure~\ref{fig_Pi_vs_Omega} demonstrates that for simple systems changes of the shear
polarization~$\Pi(t)$ can go hand in hand with those of the spatial frequency
contents~$\Omega(t)$. This establishes that shear suppression constitutes the mechanism by
which quantum dynamics conforms with interference in phase space.

Interestingly, both measures single out special states: those states for which the values
of~$\Pi$ and~$\Omega$ deviate from the typical saturated system states' values.  In the
case of weakly excited single well bound state systems the special states happen to be
partial-revival states~\cite{Averbukh_PLA89,Robinett_PR04}, see
Fig.~\ref{fig_Soft_Tableau}. Some details of Fig.~\ref{fig_Pi_vs_Omega} can be understood
from the observation that even partial-revival states feature more symmetric interference
patterns, which lowers their frequency contents~$\Omega$, when compared with odd
partial-revival states.

We emphasize that~$\Pi$ and $\Omega$ can measure aspects of
\begin{figure}[H]
  \centering 
\includegraphics[width=0.99\columnwidth]{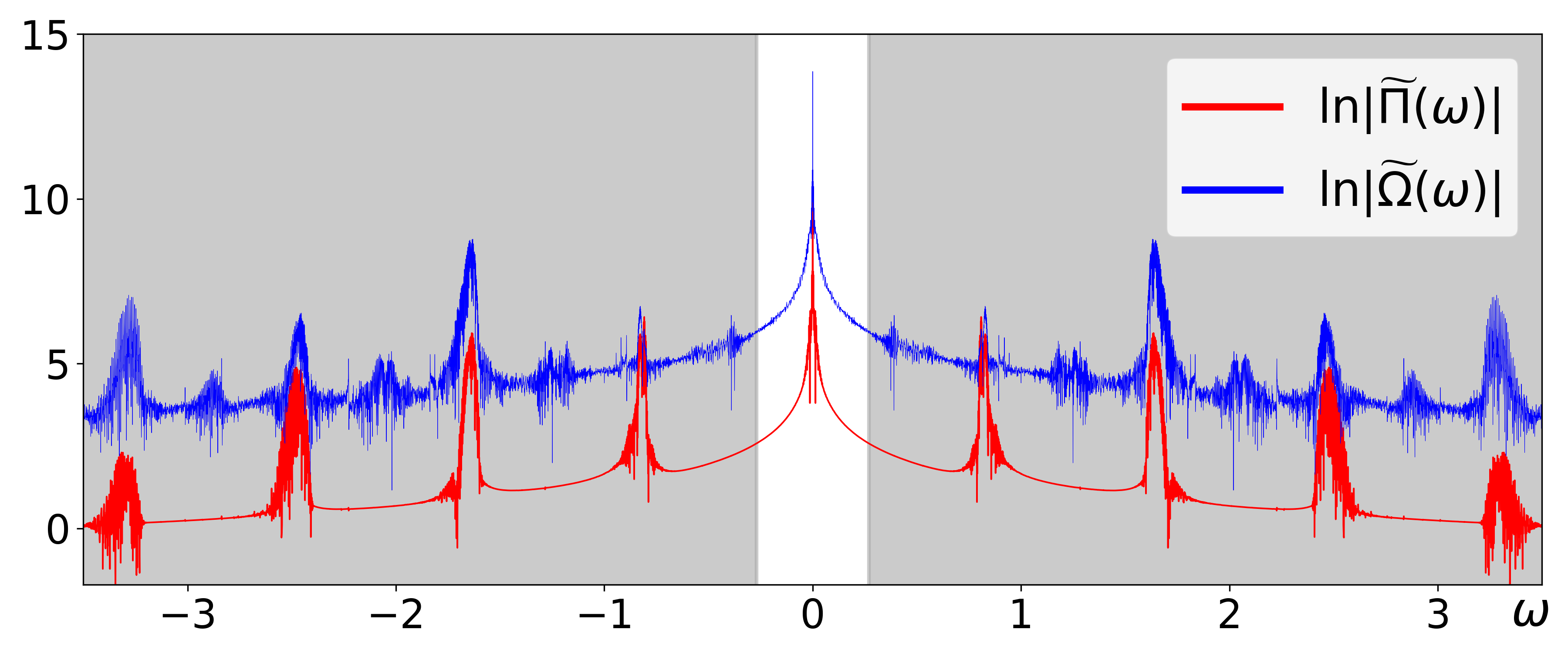}
  \caption{\emphCaption{Fourier spectra $\widetilde \Pi(\omega)$ and
      $\widetilde \Omega(\omega)$ of the time series $\Pi(t)$ and $\Omega(t)$ in
      Fig.~\ref{fig_Pi_vs_Omega}}. Note that $\Pi(t)$ provides a smoother spectrum than
    $\Omega(t)$. Cutting out $\widetilde \Pi$'s central band ($\widetilde \Pi_0$ within white
    corridor) allows us to smooth $\Pi(t)$; see Fig.~\ref{fig_Soft_Tableau}.
    \label{fig_PiTilde_vs_OmegaTilde}}
\end{figure}
\begin{figure}[H]
  \centering
  \includegraphics[width=0.99\columnwidth,height=0.55\columnwidth]{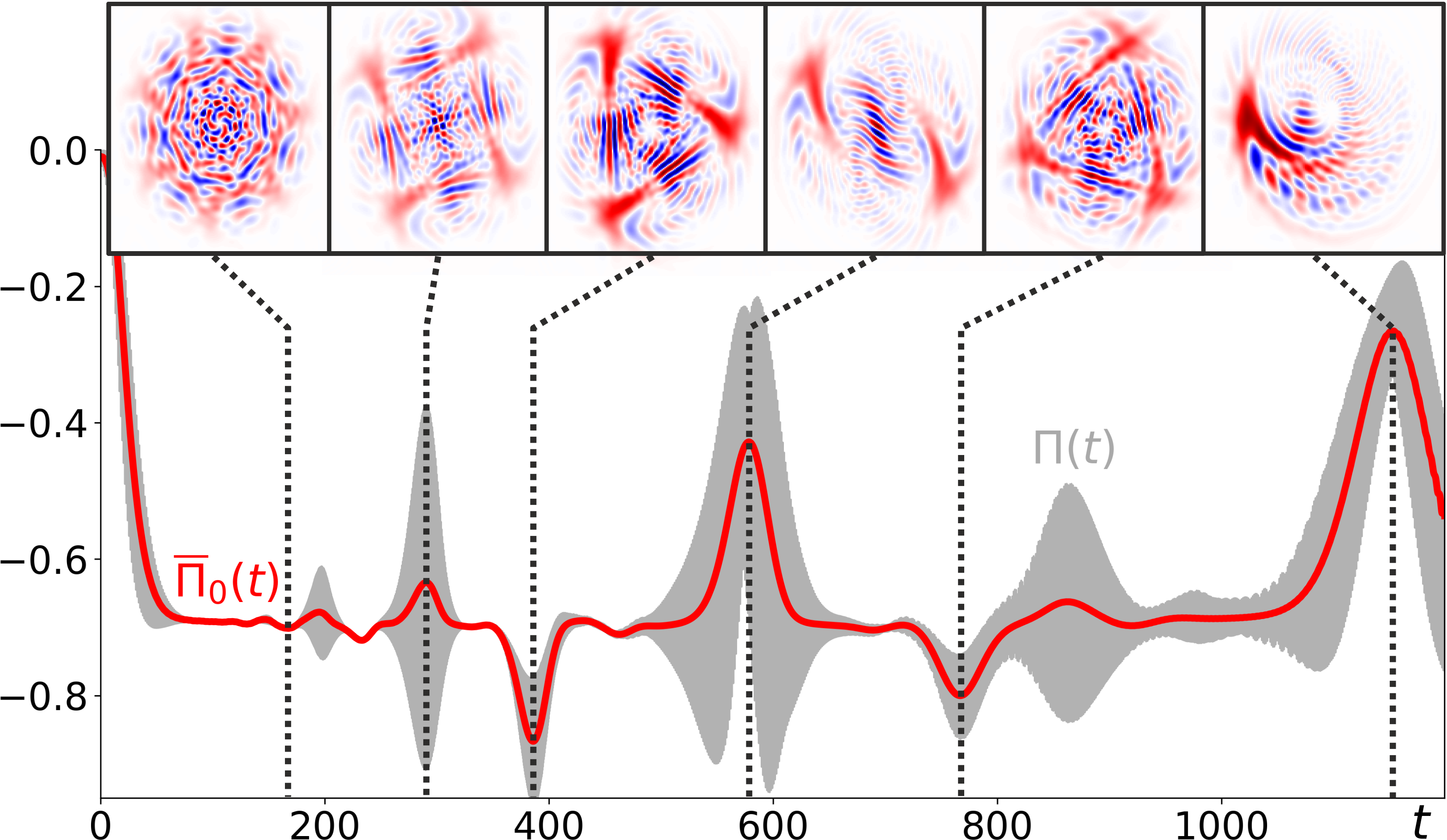}
  \caption{\emphCaption{Smoothed $\Pi(t)$ picks out special states.}  The inverse Fourier
    transform of the central band~$\widetilde{\Pi}_0(\omega)$ (highlighted in
    Fig.~\ref{fig_PiTilde_vs_OmegaTilde}) yields a (thick red) smoothed
    curve~$\overline{\Pi}_0(t)$ of~$\Pi(t)$ (gray curve). Deviations
    of~$\overline{\Pi}_0(t)$ from the settled value ($\approx -0.7$) singles out
    ``unsaturated'' special states: the evolution shows an approximate recurrence of the
    initial state at time~$T \approx 1154$. Pronounced peaks and troughs at intermediate
    times identify fractional revival states\protect\refAppendix{, see}{Averbukh_PLA89}{
      and Appendix~}{sec:Appendix_SpecialStates}{,}{} with special $n$-fold symmetries.
   \label{fig_Soft_Tableau}}
\end{figure}

\noindent
the dyna\-mics very differently from each other, see\refAppendix{
  Appendix~}{}{}{sec:Appendix_2_different}{}{ for an illustration.}

The spectrum~$\widetilde \Pi(\omega)$ of~$\Pi(t)$ is smoother than the
spectrum~$\widetilde \Omega(\omega)$ of~$\Omega(t)$, see
Fig.~\ref{fig_PiTilde_vs_OmegaTilde}. Valuable information is more easily accessible
through~$\Pi$ than~$\Omega$ since this smoothness allows us to cut out frequency bands
without sensitive dependence on the cut location. Additionally,~$\Pi$ provides information
more readily than the typically used wave function
overlap~${\cal P}(t) = |\langle \Psi_0 | \Psi (t) \rangle|^2$. This is because
${\cal P}(t)$ depends sensitively on its initial state~$\Psi_0$, but 
also because the spectrum of ${\cal P}(t)$ is noisier and does not have a
central peak that provides accessible information in the manner that $\widetilde \Pi_0$
does, see\refAppendix{ Appendix~}{}{}{sec:Compare_Pi_P}{ for details}{.}

Our approach can be applied to a wide range of systems including
Kerr systems~\cite{Oliva_Kerr_18}, driven and dissipative systems~\cite{Friedman_17},
higher-dimensional continuous systems~\cite{Wigner_PR32} and discrete spin
systems~\cite{Klimov_RMex02,Tilma_PRL16}.

\emph{To conclude,} quantum dynamics in \ps can be effectively viscous; we have traced
this back to the behaviour of the quantum corrections in Wigner's phase space
current~$\VEC J$. Quantum suppression of classical shear generates shear polarization
patterns that characterize the difference between quantum and classical \ps
dynamics. $\VEC J$'s viscosity limits the fineness of structures formed in quantum \ps
dynamics. The quantification of shear polarization patterns using~$\Pi(t)$ provides new
insight into the character of quantum \ps dynamics. Additionally, studying the time series
of~$\Pi$ we find that it sensitively displays features of the dynamics, picks out special
quantum states, does not rely on arbitrarily chosen reference states, can be frequency
filtered and provides information on the dynamics in a robust way.

For the study of the dynamics of continuous quantum systems we expect that the shear suppression
polarization $\Pi(t)$ will prove to be a valuable alternative to the wave function overlap
probability~${\cal P}(t)$.

{\bf Acknowledgements} O.S. thanks Michael Berry for his encouragement to pursue this
research and Alan McCall and Martin Hardcastle for their careful reading of the
manuscript.

\email{Ole.Steuernagel@gmail.com}



%

\clearpage

\begin{widetext}
\newpage

\setcounter{page}{1}
\setcounter{section}{0}
\renewcommand{\theequation}{\thesection~\arabic{equation}}
\renewcommand{\emphLabel}[1]{\textcolor{black}{\textbf{{#1}}}}

\begin{center}
{\large {Dynamic shear suppression in quantum \ps} -- Supplemental Material}
\\
Maxime Oliva and Ole Steuernagel
\end{center}

 \section{Fast oscillations and frequency filtering\label{sec:Appendix_Frequency_Filtering}}

\begin{figure*}[h]
\centering
\includegraphics[width=155mm,height=80mm,angle=0]{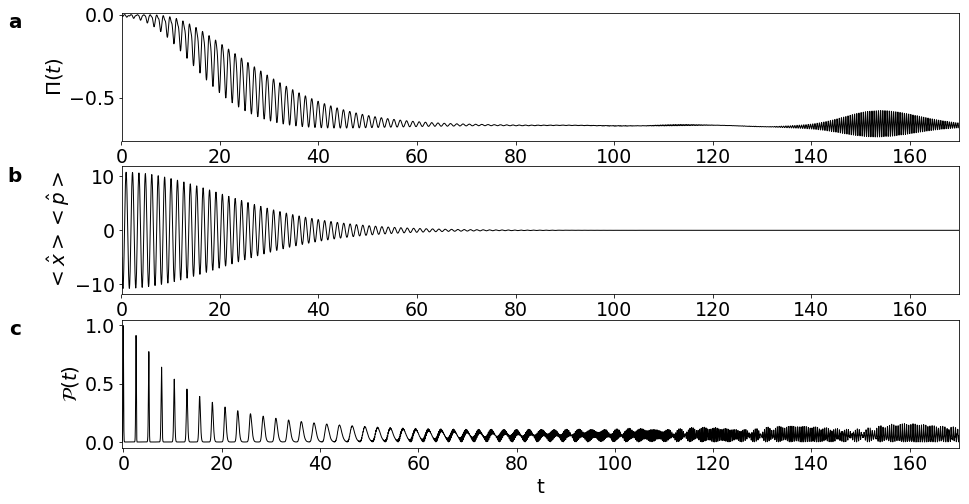}
  \caption{\emphCaption{Frequencies for \textcolor{black}{potential~$V_{\sf V}$ with parameters of
      Fig.~\ref{fig:GlobalShear}~\emphLabelbracketed{b}}.} Panel~\emphLabel{a},~$\Pi$ contains high frequency
    components at twice the frequency of the center-of-mass oscillation of the distribution~$W$, as
    evidenced by comparison with panel~\emphLabel{b} showing
    $\langle \hat x \rangle \langle \hat p \rangle$ and panel~\emphLabel{c} showing the overlap
    probability~${\cal P}(t)$. For times greater than 130 the dispersion of the state into a
    distribution with several humps creates higher harmonics frequency side-bands~$\widetilde \Pi_n$,
    compare Figs.~\ref{fig:Progressivefiltering_SOFT}
    and~\ref{fig_Soft_Tableau}~\emphLabelbracketed{b}.
    \label{fig:frequency_SOFT}}
\end{figure*}

\begin{figure*}[h]
\centering
  \includegraphics[width=155mm,height=80mm,angle=0]{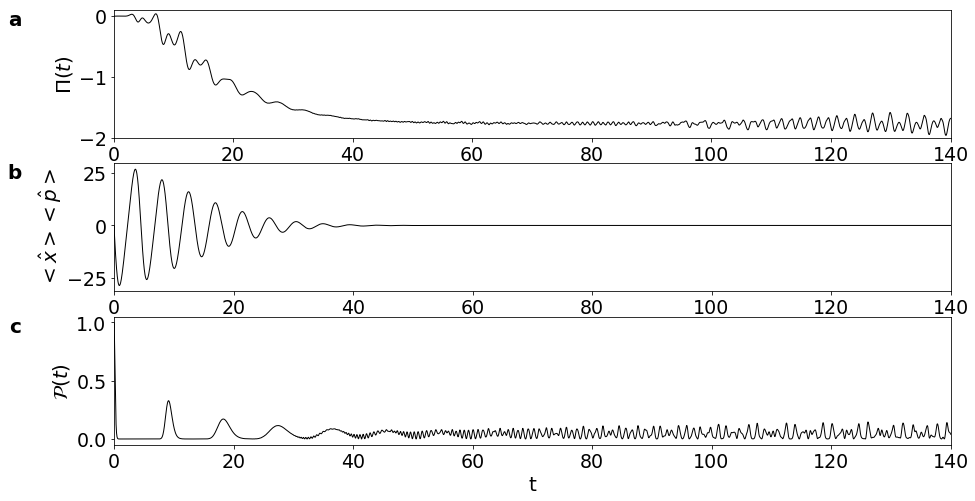}
  \caption{\emphCaption{Frequencies \textcolor{black}{for potential~$V_{\sf U}$ with parameters of
      Fig.~\ref{fig:GlobalShear}~\emphLabelbracketed{a}}.}
    Panel~\emphLabel{a},~$\Pi$ contains high frequency components at twice the frequency of the
    center-of-mass oscillation of the distribution~$W$, as evidenced by comparison with
    panel~\emphLabel{b} showing $\langle \hat x \rangle \langle \hat p \rangle$ and
    panel~\emphLabel{c} showing the overlap probability~${\cal P}(t)$, compare
    Fig.~\ref{fig_Hard_Tableau}~\emphLabel{b}.
    \label{fig:frequency_x4}}
\end{figure*}
\end{widetext}
\newpage

\onecolumngrid
\section{Comparison of shear polarization~$\Pi$ with overlap~$\cal P$\label{sec:Compare_Pi_P}}

In Figs.~\ref{fig:Progressivefiltering_SOFT}-\ref{fig:Filtering_A_x4} we apply essentially
the same filtering procedure as was used to generate
Fig.~\ref{fig_Soft_Tableau}.  Comparing the respective spectra explains
differences between~$\widetilde \Pi(\omega)$ versus~$\widetilde {\cal P}(\omega)$,
and thus~$\overline \Pi(t)$ versus~$\overline {\cal P}(t)$:

The~$\overline {\cal P}_n(t)$-curves show that characterization of the behaviour of the
system's dynamics is easier to achieve using~$\overline{\Pi}_n(t)$ than~$\overline {\cal P}_n(t)$,
compare Fig.~\ref{fig:Progressivefiltering_SOFT}~\emphLabel{b}
(and~\ref{fig_Soft_Tableau})
with~\ref{fig:Filtering_A_SOFT_Appendix}~\emphLabel{b}, or
Fig.~\ref{fig:Progressivefiltering_x4}~\emphLabel{b}
(and~\ref{fig_Hard_Tableau}~\emphLabel{e})
with~\ref{fig:Filtering_A_x4}~\emphLabel{b}.

The reason for this observation is the presence of structure in the zero-frequency band
$\widetilde \Pi_0(\omega)$, highlighted in Figs.~\ref{fig_PiTilde_vs_OmegaTilde}
and~\ref{fig_Hard_Tableau}~\emphLabel{d}; this structure can provide us with a useful
smoothed signal~$\Pi_0(t)$, see Figs.~\ref{fig_Soft_Tableau}
and~\ref{fig_Hard_Tableau}~\emphLabel{e}. In contrast to~ $\widetilde
\Pi_0(\omega)$,~$\widetilde {\cal P}_0(\omega)$ is mostly concentrated into a single isolated
peak, see Figs.~\ref{fig:Filtering_A_SOFT_Appendix}~\emphLabel{a}
or~\ref{fig:Filtering_A_x4}~\emphLabel{a}, and, as a
consequence,~$\overline {\cal P}_0(t)$ flatlines, see
Figs.~\ref{fig:Filtering_A_SOFT_Appendix}~\emphLabel{b}
or~\ref{fig:Filtering_A_x4}~\emphLabel{b}.

Additionally, the weights of the spectral bands~$\widetilde \Pi_n(\omega)$ drop with
increasing band index~$n$, see Figs.~\ref{fig:Progressivefiltering_SOFT}~\emphLabel{a}
(and \ref{fig_Soft_Tableau}), or Figs.~\ref{fig:Progressivefiltering_x4}~\emphLabel{a}
(and \ref{fig_Hard_Tableau}~\emphLabel{b}). Higher order bands can be truncated without
losing too much information.  In contrast, the weights of the spectral
bands~$\widetilde {\cal P}_n(\omega)$, see
Figs.~\ref{fig:Filtering_A_SOFT_Appendix}~\emphLabel{a}
or~\ref{fig:Filtering_A_x4}~\emphLabel{a}, remain similar across several frequency
bands~$n$. For useful information, bands with high index~$n$ have to be retained. Their
associated time-signal therefore suffers from complexity-overload, contrast
Fig.~\ref{fig:Filtering_A_SOFT_Appendix}~\emphLabel{b}
with~\ref{fig:Progressivefiltering_SOFT}~\emphLabel{b},
or~\ref{fig:Filtering_A_x4}~\emphLabel{b}
with~\ref{fig:Progressivefiltering_x4}~\emphLabel{b}.

The signals~$\overline {\cal P}_n(t)$ display spurious negativities, see
Figs.~\ref{fig:Filtering_A_SOFT_Appendix}~\emphLabel{b}
or~\ref{fig:Filtering_A_x4}~\emphLabel{b}, because they are filtered before being
back-transformed. The probabilities~${\cal P}(t)$ are of course positive at all times, see
Figs.~\ref{fig:frequency_SOFT}~\emphLabel{c} and~\ref{fig:frequency_x4}~\emphLabel{c}.

\subsection{Case of soft potential~$V_{\sf V}$\label{subsec:Compare_Pi_P_soft}}
\twocolumngrid
\renewcommand{\emphLabel}[1]{\textcolor{black}{\textbf{{#1}}}}
${}^{}$
\vspace{0.6cm}

\begin{figure}[h]
\centering
\includegraphics[width=\squeezedWidth,height=\squeezedHeight]{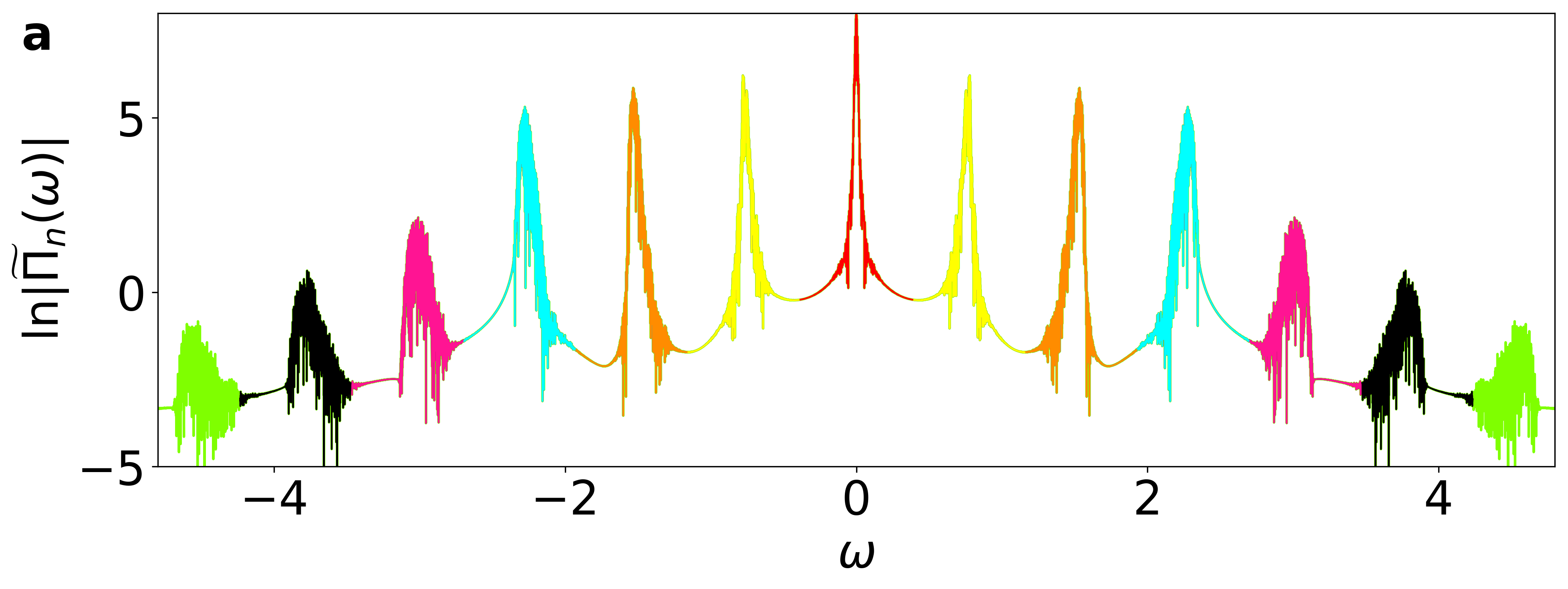}
  \\
  \includegraphics[width=\squeezedWidth,height=\squeezedHeight]{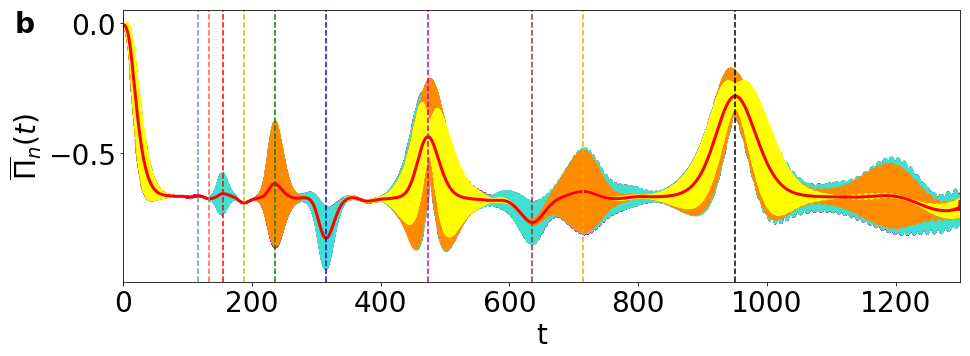}
  \caption{\emphCaption{Smoothing of~$\Pi(t)$ through frequency filtering.}  For the
    parameters of Fig.~\ref{fig:GlobalShear} \emphLabelbracketed{b} with
    potential~$V_{\sf V}$: Panel~\emphLabel{a},~harmonic frequency bands of the Fourier
    image~$\widetilde \Pi(\omega)$ of~$\Pi(t)$ are color-labeled. We progressively remove
    higher harmonics and retain only the $n$ lower order bands~$\widetilde \Pi_n$ grouped
    in pairs around the central band~$\widetilde \Pi_0$ at zero (red
    color).~\emphLabel{b}, when back-transforming~$\widetilde \Pi_n$ we arrive at smoothed
    curves $\overline \Pi_n$ of~$\Pi$, color-labeled by the highest retained frequency
    band in~\emphLabel{a} above.  Panel~\emphLabel{b} shows that
    the~$\overline \Pi_n$-curves pick out special states of the corresponding recurrence
    order~$n$ at recurrence times~$T/n$ and their multiples, compare
    Fig.~\ref{fig_Soft_Tableau}.
    \label{fig:Progressivefiltering_SOFT}}
\end{figure}
\newpage
\begin{figure}[h]
\centering
\vspace{1.3cm}
\includegraphics[width=\squeezedWidth,height=\squeezedHeight]{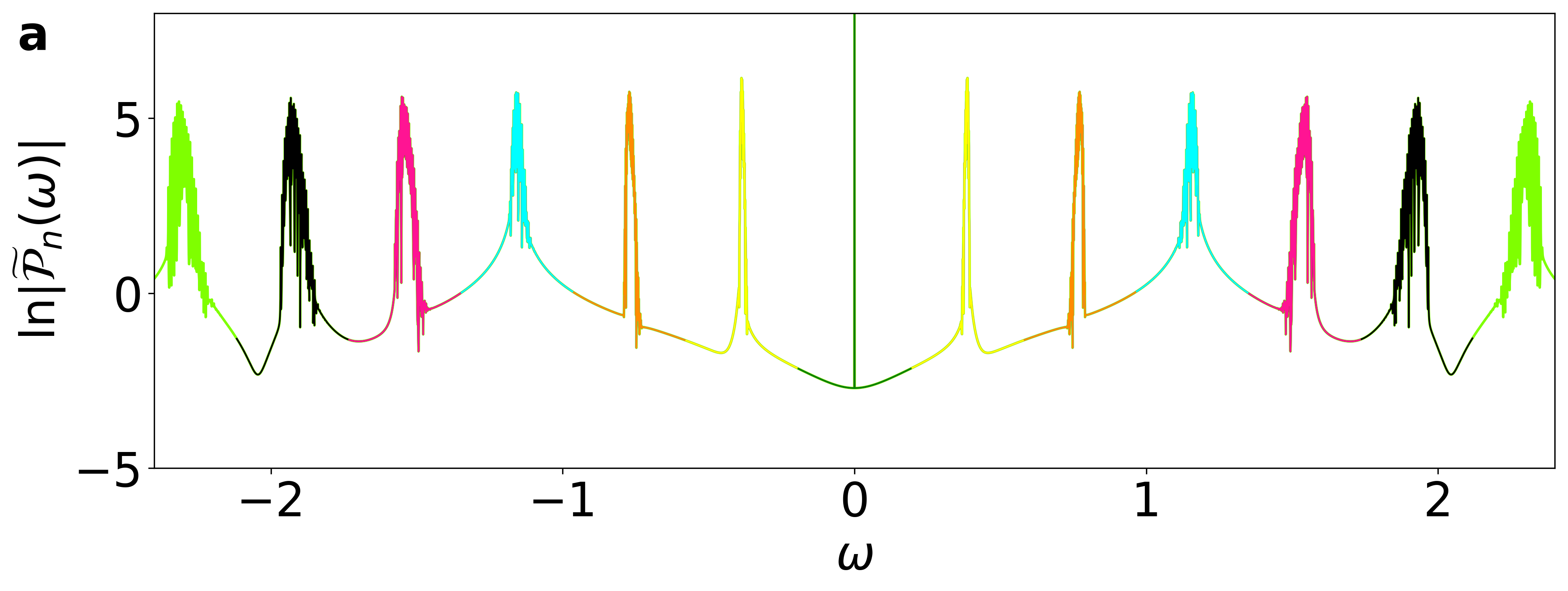}
  \\
  \includegraphics[width=\squeezedWidth,height=\squeezedHeight]{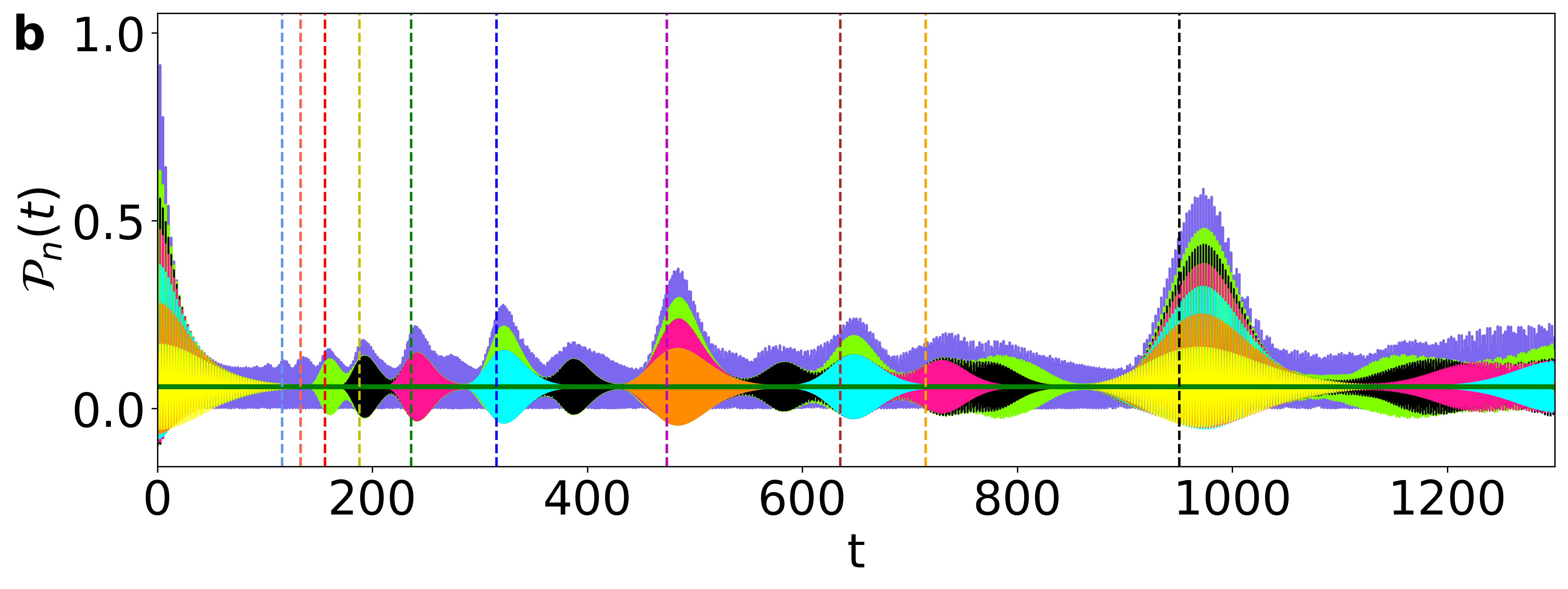}
  \caption{\emphCaption{Smoothing of~${\cal P}(t)$ through frequency filtering.}  For the
    parameters of Fig.~\ref{fig:GlobalShear} \emphLabelbracketed{b} with
    potential~$V_{\sf V}$: Panel~\emphLabel{a}, shows~harmonic
    frequency bands of the Fourier image~$\widetilde {\cal P}(\omega)$ of~${\cal P}(t)$ and
    panel~\emphLabel{b}, the back-transformed images of~$\widetilde {\cal P}_n$, yielding
    smoothed curves $\overline {\cal P}_n$ of~${\cal P}$, compare
    Figs.~\ref{fig_Soft_Tableau},~\ref{fig:Progressivefiltering_SOFT}
    and~\ref{fig_Hard_Tableau}.~\emphLabel{b}, when
    back-transforming groups~$\widetilde {\cal P}_n(\omega)$ we arrive at smoothed curves
    $\overline {\cal P}_n(t)$ of~${\cal P}(t)$, color-labeled by the highest retained
    frequency band from~\emphLabel{a} above; $\overline {\cal P}_0(t)$ flatlines. 
    The dashed lines of panel~\emphLabel{b} have been carried
    over from Fig.~\ref{fig_Soft_Tableau}. Their slight time offset is due
    to the fact that $\overline {\cal P}_n$ only measures the overlap with the initial 
    state, whereas $\overline \Pi$ is a global
    measure.
    \label{fig:Filtering_A_SOFT_Appendix}
  }
\end{figure}

\clearpage
\onecolumngrid
\subsection{Case of hard potential~$V_{\sf U}$\label{subsec:Compare_Pi_P_hard}}
\twocolumngrid

\begin{figure}[h]
\centering
\includegraphics[width=\squeezedWidth,height=\squeezedHeight]{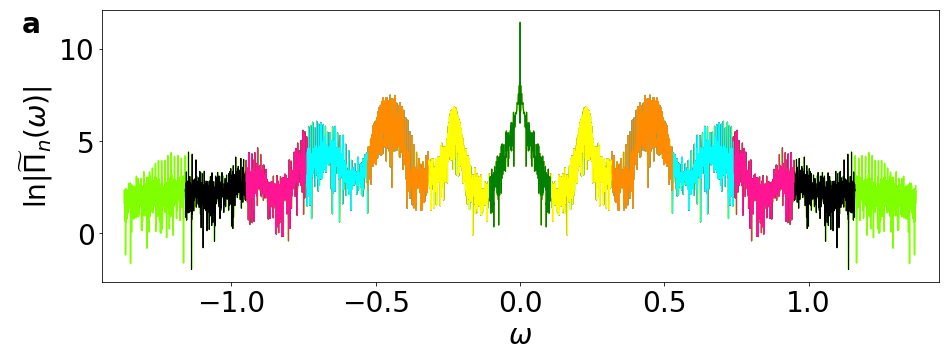}
  \\
  \includegraphics[width=\squeezedWidth,height=\squeezedHeight]{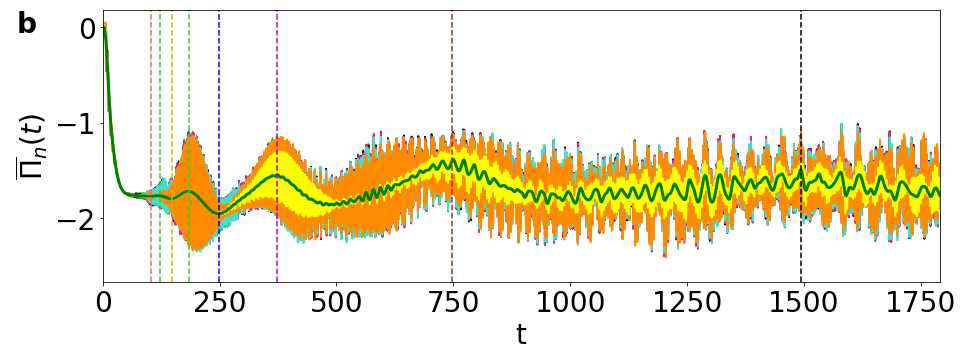}
  \caption{\emphCaption{Smoothing of~$\Pi(t)$ through frequency filtering.}
    For the
    parameters of Fig.~\ref{fig:GlobalShear} \emphLabelbracketed{a} with hard
    potential~$V_{\sf U}$: Panel~\emphLabel{a},~harmonic
      frequency bands of the Fourier image~$\widetilde \Pi(\omega)$ of~$\Pi(t)$ are
      color-labeled. We progressively remove higher harmonics and retain only the $n$
      lower order bands~$\widetilde \Pi_n$ grouped in pairs around the central
      band~$\widetilde \Pi_0$ at zero (dark green color).~\emphLabel{b}, when
      back-transforming~$\widetilde \Pi_n$ we arrive at smoothed curves $\overline \Pi_n$
      of~$\Pi$, color-labeled by the highest retained frequency band in~\emphLabel{a}
      above.  Panel~\emphLabel{b} shows that the~$\overline \Pi_n$-curves pick out special
      states of the corresponding recurrence order~$n$ at recurrence times~$T/n$ and their
      multiples, compare Fig.~\ref{fig_Hard_Tableau}.
      \label{fig:Progressivefiltering_x4}}
\end{figure}

\newpage

\begin{figure}[h]
\centering
  \includegraphics[width=87mm,height=33mm]{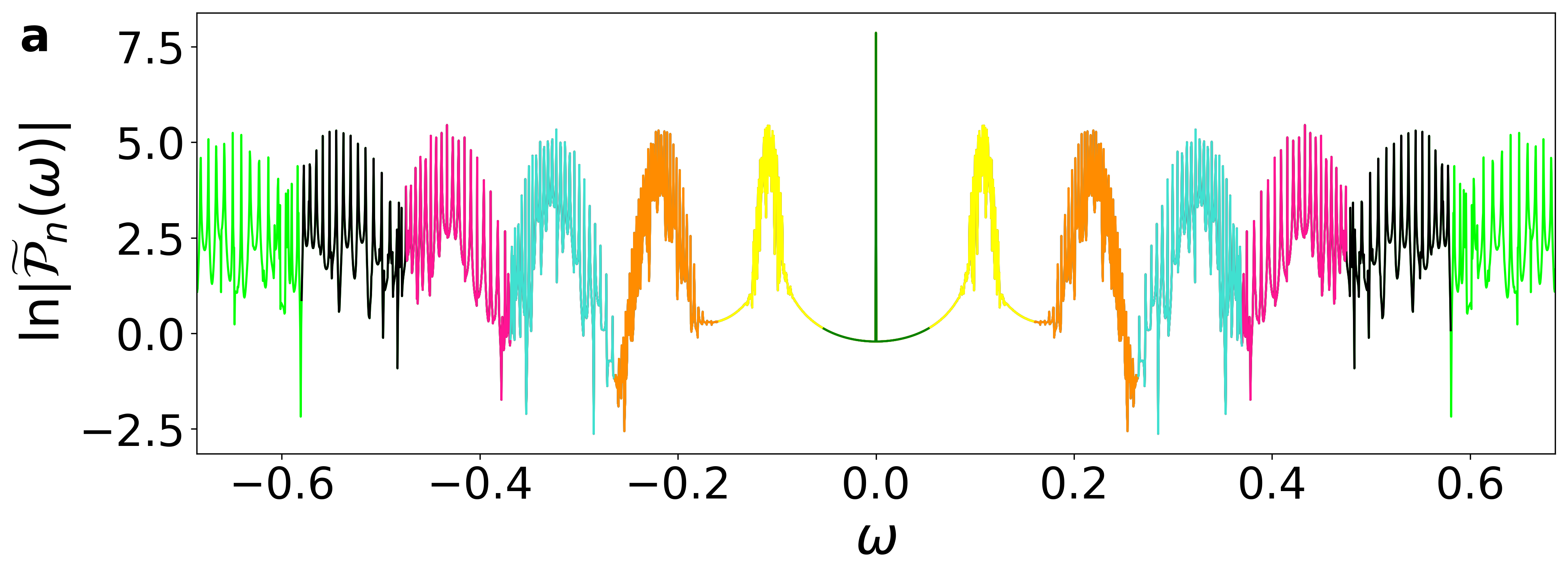}
  \\
  \includegraphics[width=87mm,height=33mm]{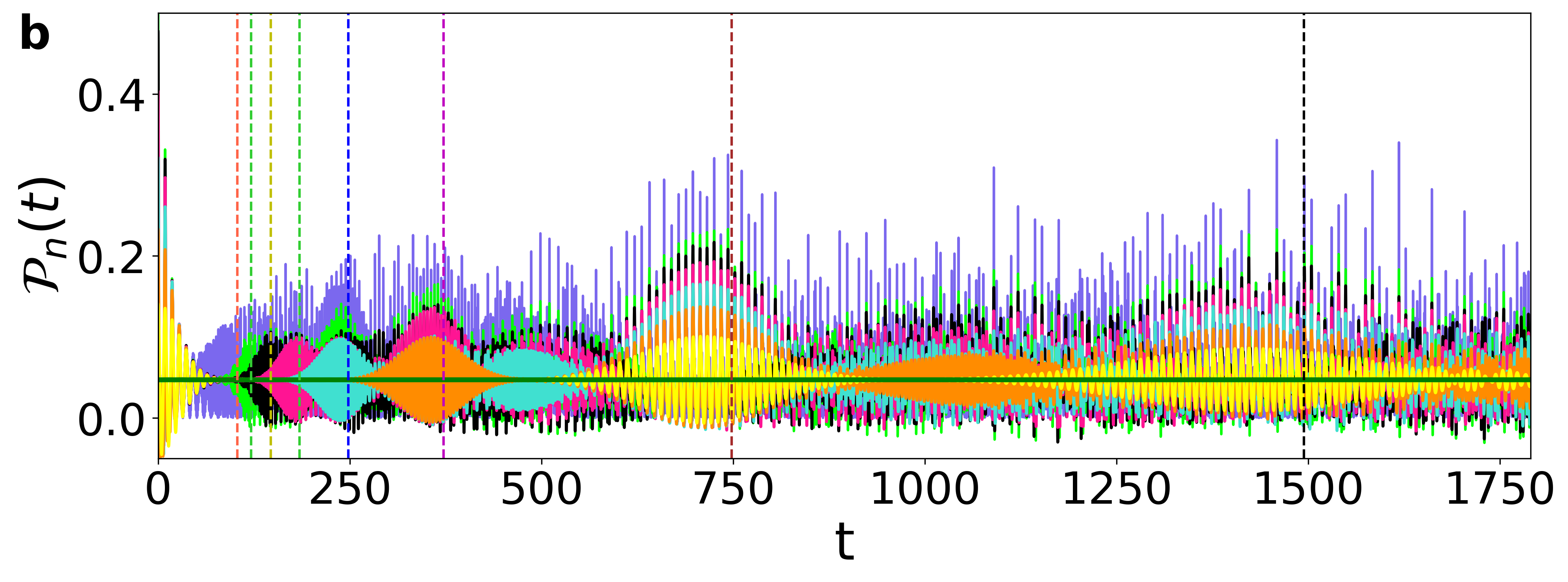}
  \caption{\emphCaption{Smoothing of~${\cal P}(t)$ through frequency filtering.} For the
    parameters of Fig.~\ref{fig:GlobalShear} \emphLabelbracketed{a} with hard
    potential~$V_{\sf U}$: Panel \emphLabel{a},~harmonic frequency bands of the Fourier
    image~$\widetilde {\cal P}(\omega)$ of~${\cal P}(t)$ are color-labeled. We
    progressively remove higher harmonic bands and retain the
    group~$\widetilde {\cal P}_n(\omega)$ of all $n$ lower order bands. The smal\-lest
    such group is the central band~$\widetilde {\cal P}_0(\omega)$ around zero (dark green
    color).~\emphLabel{b}, when back-transforming groups~$\widetilde {\cal P}_n(\omega)$
    we arrive at smoothed curves $\overline {\cal P}_n(t)$ of~${\cal P}(t)$,
    color-labeled by the highest retained frequency band from~\emphLabel{a} above;
    $\overline {\cal P}_0(t)$ flatlines.  The dashed lines of panel~\emphLabel{b} have
    been carried over from Fig.~\ref{fig_Hard_Tableau}~\emphLabel{e}. Their slight time
    offset is due to the fact that $\overline {\cal P}_n$ only measures the overlap with
    the initial state, whereas $\overline \Pi$ is a global measure.}
    \label{fig:Filtering_A_x4}
\end{figure}

\newpage
\begin{widetext}
\section{Identification of special states\label{sec:Appendix_SpecialStates}}

The identification of special states, performed in~Fig.~\ref{fig_Soft_Tableau} for the
soft potential~$V_{\sf V} = {31 x^2}/{10} - { x^4}/{81}$, can also be performed for the
hard potential case, $V_{\sf U} = x^4 /500$. This is illustrated in
Fig.~\ref{fig_Hard_Tableau} below:

\begin{figure*}[h]
\centering
\includegraphics[width=175mm,height=150mm,angle=0]{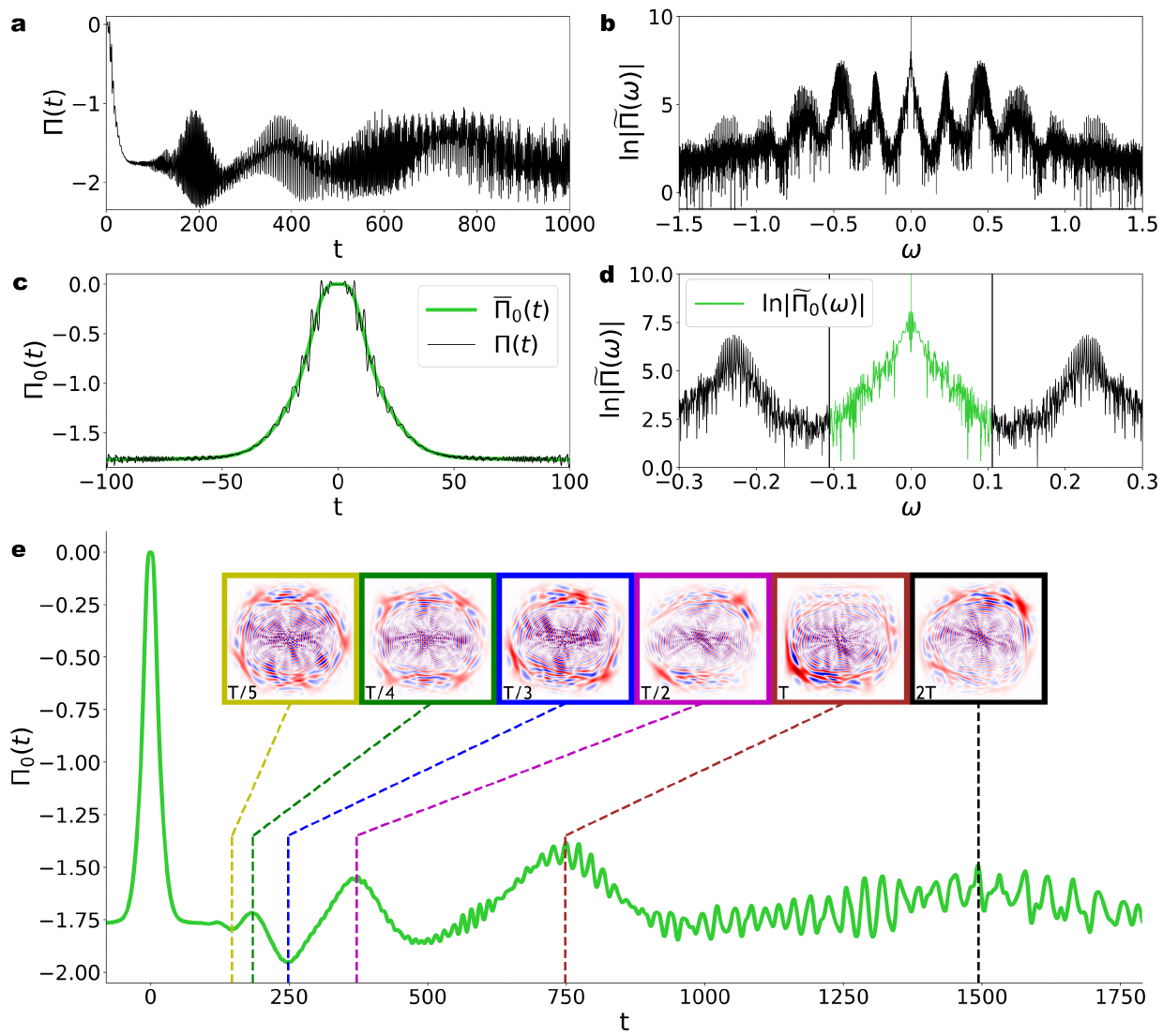}
\caption{\emphCaption{Smoothed $\Pi(t)$, picks out special states.} For the
  \textcolor{black}{same hard potential~$V_{\sf U}$ and state, as in
    Fig.~\ref{fig:GlobalShear}~\emphLabelbracketed{a}}: \emphLabel{a},~$\Pi(t)$ contains high
  frequency components, \emphLabel{b}, which are grouped into harmonic
  bands. \emphLabel{d}, the central band (highlighted in green) is cut out and its inverse
  Fourier transform gives, \emphLabel{c}~and~\emphLabel{e}, smoothed
  profiles~$\overline{\Pi}_0(t)$ of~$\Pi(t)$ in~\emphLabel{a}. \emphLabel{c}, the close-up
  of~${\Pi(t)}$ and~$\overline{\Pi}_0(t)$ near~$t=0$ shows the symmetry with respect to
  the ``most unsettled'' initial state, compare Fig.~\ref{fig:GlobalShear} and main
  text. \emphLabel{e},~when~$\Pi$'s value deviates most from the longtime average we find
  that the evolution has led to an approximate recurrence of the initial state at
  time~$T \approx 750$.  One observes several pronounced peaks and troughs at intermediate
  times where fractional revival states~\cite{Averbukh_PLA89} with special $n$-fold
  symmetries are found.
  \label{fig_Hard_Tableau}}
\end{figure*}

\newpage
\section{Some properties of~$W$ and $\VEC J$ \label{sec_W_Special}}

Wigner's \ps quantum distribution~\cite{Wigner_PR32,Hillery_PR84}
\begin{equation}\label{eq:W}
  W_\varrho(x,p,t) \equiv \frac{1}{\pi \hbar} \int_{-\infty}^{\infty}
  dy \, e^{-\frac{2i}{\hbar} p y} 
  \langle x+y| \hat \varrho(t) | x-y \rangle 
\end{equation}
is real-valued and normalized $ \iint_{-\infty}^{\infty} dx \; dp \; W(x,p,t) = 1 $.
Since $W_\varrho$ is $\varrho$'s Fourier transform, $W_\varrho$ represents exactly the
same information about a quantum system as the density matrix~$\varrho$ itself, here we
therefore only investigate~$W$.

By construction $W$ is nonlocal in~$y$ and subject to Fourier-limits; the same applies to
the components of~$\VEC{J}$:

 Its position component is
\begin{flalign}
  J_x  = \frac{p}{M\pi\hbar}\int
  dy\; \varrho(x-y,x+y,t)e^{\frac{2i}{\hbar}py} = \frac{p}{M} W, &
  \label{eq:Wigner_Current_Jx}
\end{flalign}
and its momentum component is
\begin{flalign}
 J_p  = 
  -\frac{1}{\pi\hbar}\int
  dy\; \left[\frac{V(x+y)-V(x-y)}{2y}\right]
  \varrho(x-y,x+y,t) e^{\frac{2i}{\hbar}py} .
 \label{eq:Wigner_Current_Jp_Integral_Form}
 \end{flalign}
 $W$ and its dynamics is set apart from other quantum \ps
 distributions~\cite{Hillery_PR84} as the closest quantum analog of the classical \ps
 distribution~\cite{Oliva_PhysA17,Hillery_PR84,Leibfried_PT98,Zurek_NAT01,Tilma_PRL16}.

\section{Unit-free formulation of vorticity  \label{sec_UnitsVorticity}}

To quantify the vorticity of~$\VEC{v}$ in \ps, we form
$ -\VEC{\nabla} \times \VEC{v} = \frac{\xi}{\tau} \frac{\partial}{\partial{\cal P}} v_{\cal X} -
\frac{\tau}{\xi} \frac{\partial}{\partial{\cal X}} v_{\cal P} =\sqrt{MK}
\frac{\partial}{\partial{\cal P}} v_{\cal X} - \frac{1}{\sqrt{MK}}
\frac{\partial}{\partial{\cal X}} v_{\cal P}$. This form is firstly inspired by the curl
operator and can secondly be justified by the fact that shear stress in classical fluids
is proportional to its transverse velocity derivatives, see Chap.~41, Vol.~II of
\cite{Feynman_Lectures}. In this second interpretation the minus sign arises from the
symplectic structure of \ps.

We then use the coordinate transformations
$x=(M K / \hbar^2)^{1/4} {\cal X} = \xi {\cal X}$ and $p=
\tau {\cal P}$, where $\tau = 1/(\hbar \xi)$ to switch to unit-free systems ($\hbar = 1$,
$M=1$ and $K=1$). This, e.g., maps a harmonic oscillator Hamiltonian, given in physical
units,
$\hat {\cal H}= - \frac{\hbar^2}{2M} \frac{\partial^2}{\partial{\cal X}^2} + \frac{K}{2}
{\cal X}^2$, to the isomorphic, unit-free system
$ \hat H =- \frac{1}{2} \frac{\partial^2}{\partial{x}^2} + \frac{x^2}{2}$ (with circular
\ps trajectories~\cite{Kakofengitis_EPJP17}); applied to~$\VEC{v}$ it yields
Eq.~(\ref{eq:_ClassicalShear}).

\section{Structure formation and coherences  \label{sec_StructureFormation}}

We study one-dimensional systems, continuous in position~$x$ and momentum~$p$, the
generalization to higher dimensional continuous systems is
straightforward~\cite{Wigner_PR32}.

It is known that forced oscillators with Hamiltonians of the
form~$\hat {\cal H}= - \alpha(t) \frac{\hbar^2}{2M} \frac{\partial^2}{\partial{\cal X}^2}
+ \gamma(t) \hat{\cal X} + \beta(t) \hat{\cal X}^2$, with real functions~$\alpha, \beta$
and~$\gamma$, have classical solutions in the sense that trajectories for the transport of
their Wigner distribution can be given~\cite{Takabayasi_PTP54,Oliva_PhysA17}. This implies
that systems described by such Hamiltonians cannot create or remove quantum coherences,
or, what is the same~\cite{Feynman_NegEssay87,Leibfried_PT98}, they cannot create or
remove~\cite{Kakofengitis_PRA17,Oliva_PhysA17} Wigner distribution negativities.

Such systems can stretch and otherwise deform Wigner distributions through classical
dynamics~\cite{Kurtsiefer_NAT97}. But, such changes do not concern us here since they do
not create or modify coherences or negativities of the quantum state. The measures we
devise here are designed to ignore
classical-only~\cite{Kakofengitis_PRA17,Oliva_PhysA17,Takabayasi_PTP54} state
transformations.

\section{On the term ``viscosity''  \label{sec_Viscosity}}

The fact that quantum dynamics does not generically allow for structure formation in phase
space below Zurek's scale~$a_Z$ is due to the fact that the quantum terms in~$\bf J$
somewhat suppress its shear.

We do not mean to imply that the dynamics is dissipative; even in classical, creeping
Stokes flow the fluid's dynamics is approximately reversible although its is dominated by
its viscous behaviour.

Also, the quantum dynamics allows for fractional and full revivals, in this case
structures are reversibly removed in \ps and the system becomes dynamically
``unsettled''.

Partly, this is poorly captured by the term viscous, it should only be understood as a
superficial description, yet we use the term viscous since no better term seems to
exist.

\newpage

 \section{Derivation of inequality for frequency content $\Omega$ in
   Eq.~(\ref{eq:Omega_W_FrequencyContent})\label{sec:Appendix_OmegaLimit}}
  For states bound in a single well one can estimate~$W$'s extent in phase space according
  to Zurek's arguments~\cite{Zurek_NAT01}:
  \\
  assuming that~$W$ is spread over a spatial distance~$L$ and momentum distance~$P$ and is
  in its entirety structured at the associated Zurek scale as given by
  Eq.~(\ref{eq:aZurek}), it follows that its frequency spread in $k_x$ is roughly
  $2K_x=2P/\hbar$ and in $k_p$ is roughly $2K_p=2L/\hbar$; this is easily confirmed
  analytically. 
  \\
  Furthermore, assuming that~$\overset{\approx}{W}(k_x,k_p)$ is confined to an ellipse
  with semi-axes $K_x$ and~$K_p$ in \ps (rather than the exact energy contour associated
  with the Hamiltonian~$H$) we can bound $\Omega$ by assuming that all the weight
  of~$\overset{\approx}{W}(k_x,k_p)$ is concentrated on this elliptical rim alone. This
  yields the desired inequality
\begin{flalign}
  \Omega = \frac{\iint dk_x dk_p \; |\overset{\approx}{W}(k_x,k_p,t) \; k_x k_p|}{\iint
    dk_x dk_p \; |\overset{\approx}{W}(k_x,k_p,t)|} < \oint d \varphi | K_x \cos(\varphi)
  K_p \sin(\varphi)| = \Omega_\text{max} = 2 K_x K_p = \frac{8 \pi^2}{a_Z}.
 \label{eq:Omega_Bound}
 \end{flalign}

 \section{Remarks on measures $\delta$, $\pi$ and
   $\Pi$\label{sec:Appendix_RemarksMeasures}}

 The main structural difference between a classical shear measure such
 as~$s$ and the quantum measure~$\pi$ is, that the latter is based on
 the \emph{difference} $\VEC{J}-\VEC{j} = \VEC{J}^Q$.

The reason is somewhat subtle and deserves further discussion:

Classical shear~$s$ is solely based on the Hamiltonian velocity field~$\VEC{v}$ and
therefore distinguishes between hard and soft potentials, this case distinction does in
general \emph{not} carry over to the behaviour of the current~$\VEC{j} = \rho \VEC{v}$.

After sufficiently long times quantum and
classical distributions become stretched out such that they are sharply peaked in the
direction perpendicular to their stretching.  Thus large gradients result in the direction
of~$\widehat {{\bm \nabla}}_{\!\!H}$, see Fig.~\ref{fig:Distributions},
irrespective of whether a hard or soft potential generated the distribution.

A measure such as $\delta$, see Eq.~(\ref{eq:shear_deviation}), therefore becomes dominated
by the contribution from the derivatives due to the shape of the distribution and
insensitive to the distinction between hard and soft potentials. Using the
\emph{difference} $\VEC{J}-\VEC{j}$ happens to reinstate the distinction between hard and
soft potentials.

For the interested reader this is further explored in the next Subsection.

\subsubsection*{Classical current shear polarization scales with $t^1$
\label{subsubsec:Appendix_sCaseDistinctionBreakdown}}

Consider the derivatives in
$\sigma = \partial_{\widehat {{\bm \nabla}}_{\!\!H}}  (- {\bm \nabla} \times \VEC{j})$. For sufficiently
long times and sufficiently smooth anharmonic potentials the terms containing second order
derivatives of~$\rho$ become dominant and $\iint \sigma \;dx dp \sim t$, where the
proportionality constant is positive in the case of bound systems since it is given by an
average over the Hamiltonian's gradient (which in bound systems is positive):

For long times the state stretches into a filament that grows linearly in length with
time, since in the conservative case the `angular' velocity difference between two energy
shells is constant. Additionally, the narrowing of the filament's width inversely with
time scales up curvatures, associated with the second derivatives of~$\rho$ in $\sigma$,
quadratically in time. This growth with~$t^3$, upon integration, is compensated for by the
narrowing in width~$\propto t^{-1}$ which in turn has the side effect of picking up a
linearly shrinking sample~$\propto t^{-1}$ of the (smooth) variation of~$H(x,p)$ across
energy shells. The upshot of this is that the above integral is linear in~$t$; this can be
confirmed numerically.

With the proportionality constant positive, we find, that the distinction between hard and
soft potentials disappears for classical shear measures such
as~$\sigma (\VEC{j})= \sigma (\rho \VEC{v})$ as well as their quantum
counterpart~$\sigma (\VEC{J})$.

This is why we use
\begin{flalign}
{\pi}(x,p,t;H) = W(t) \; \partial_{\widehat {{\bm \nabla}}_{\!\!H}} \delta(t;H)
\nonumber
\end{flalign}
which is a function of~$\VEC{J}^Q = \VEC{J} - \VEC{j}$ and reinstates the distinction
between hard and soft potentials.

\newpage
\section{Comparing polarization patterns for `hard' and `soft' potentials \label{sec:Appendix_LocalPolarizationPattern}}

\begin{figure*}[h]
\includegraphics[width=0.95\textwidth]{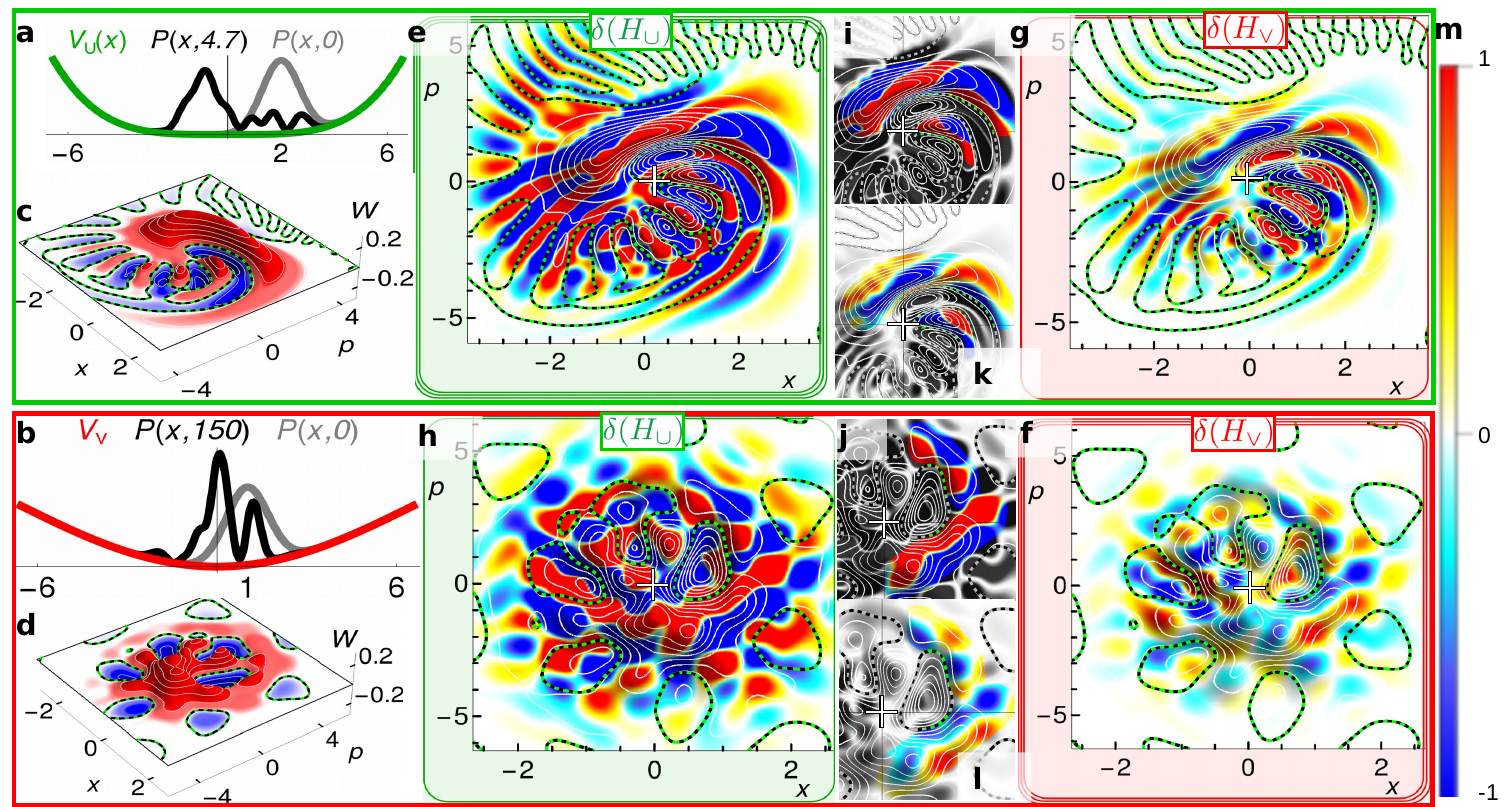}
\caption{\CO{(Color online)} \emphCaption{Polarization of deviation vorticity~$\delta$.}
  \emphLabel{a}, sketch of hard [$V_{\sf U}=x^4/16$] and, \emphLabel{b}, soft potential
  [$V_{\sf V}={31 x^2}/{10} - { x^4}/{81}$] together with probability
  distributions~$P(x,t)=|\Psi(x,t)|^2$ (black curves) of states evolved under these
  potentials from initial Gaussian states~$P(x,0)$ (grey curves) at initial center
  positions, \emphLabel{a}, $x_0=2$ and, \emphLabel{b}, $x_0=1$, respectively. The Wigner
  distributions associated with~$\Psi(x,t)$ in \emphLabel{a} and \emphLabel{b} are shown
  in \emphLabel{c} and \emphLabel{d}, respectively. \emphLabel{e}-\emphLabel{h}, contours
  of the Wigner distributions of~\emphLabel{c} or \emphLabel{d} overlaid with a color bar
  with \emphLabel{m}'s colors representing values of~tanh[$50 \; \delta(H)$],
  where~$\delta(H)$, see Eq.~(\ref{eq:shear_deviation}), is specified in the head of each
  framed panel. Small panels \emphLabel{i}-\emphLabel{l} highlight two regions where~$W>0$
  [using color while the non-highlighted background is kept gray] to highlight polarization inversion
  when the Hamiltonian is switched [see main text after
  Eq.~(\ref{eq:shear_deviation})]. For reference, the origin~$(x,p)=(0,0)$ is labeled by
  a white cross.
  \label{Appendix_fig:LocalShear}}
\end{figure*}

We note that our soft potential,~$V_{\sf V} = {31 x^2}/{10} - { x^4}/{81}$, used in
Figs.~\ref{fig:Distributions} and~\ref{fig:GlobalShear}, is formally open for large values
of~$x$ but we restrict its use to `safe' values~$|x|<10$ which allows us to ignore quantum
tunnelling out of its central well.

\section{ $\Pi(t)$ and $\Omega(t)$ can evolve
  differently from each other. \label{sec:Appendix_2_different}}

\begin{figure}[h]
\centering
\includegraphics[width=0.43\columnwidth]{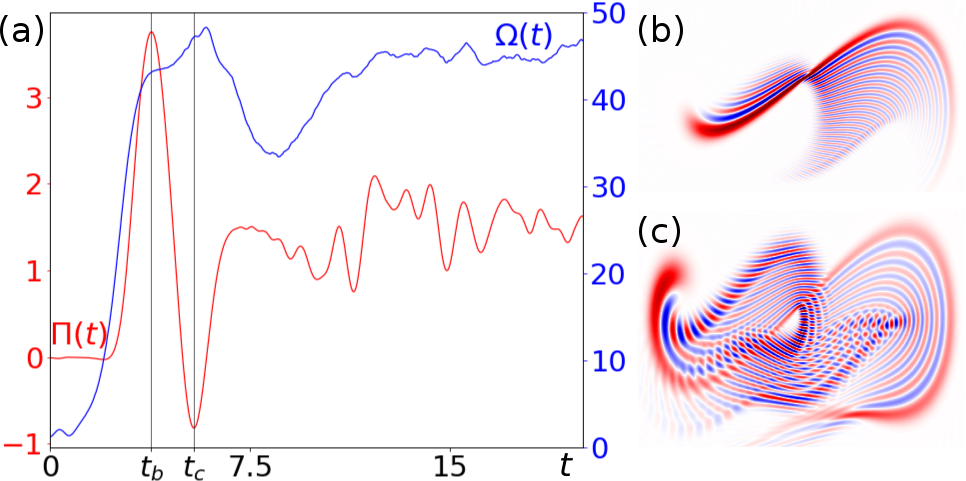}
\caption{\emphCaption{For a system more complicated than the single well potentials
    considered in the main text, shear suppression $\Pi(t)$ and frequency contents
    $\Omega(t)$ can evolve differently.} For a double well potential
  $V_{\sf W} = - \frac{x^2}{2} + \frac{x^4}{200}$ with an initial Gaussian state~$W_0$
  located at $(x_0,p_0) = (-9.9,0) $ $\Pi(t)$ shows a pronounced maximum and minimum at
  times~$t_{b} \approx 3.7$ and~$t_{c} \approx 5.4$.  $\Omega(t)$ is insensitive to this
  system's special dynamic behaviour.
  \label{fig_Seahorse}}
\end{figure}

 \clearpage
\end{widetext}




\end{document}